\documentclass[12pt]{article}
\usepackage{graphicx,amssymb,amsmath}
\usepackage{epsfig}
\usepackage{color,graphicx}
\usepackage{cite}
\usepackage{amssymb,amsmath}
\newcommand{\rc}{\nonumber\\}
\newcommand{\be}{\begin{equation}}
\newcommand{\ee}{\end{equation}}
\newcommand{\bea}{\begin{eqnarray}}
\newcommand{\eea}{\end{eqnarray}}
\newcommand{\beas}{\begin{eqnarray*}}
\newcommand{\eeas}{\end{eqnarray*}}
\newcommand{\beq}{\begin{equation}}
\newcommand{\eeq}{\end{equation}}
\newcommand{\bear}{\begin{eqnarray}}
\newcommand{\eear}{\end{eqnarray}}

\begin{document}

\begin{center}

\centerline{\Large {\bf Magnetic catalysis in flavored ABJM}}

\vspace{8mm}

\renewcommand\thefootnote{\mbox{$\fnsymbol{footnote}$}}
Niko Jokela,${}^{1}$\footnote{niko.jokela@usc.es}
Alfonso V. Ramallo,${}^{1}$\footnote{alfonso@fpaxp1.usc.es} and
Dimitrios Zoakos${}^2$\footnote{dimitrios.zoakos@fc.up.pt}

\vspace{4mm}
${}^1${\small \sl Departamento de F\'isica de Part\'iculas} \\
{\small \sl Universidade de Santiago de Compostela}\\
{\small \sl and}\\
{\small \sl Instituto Galego de F\'isica de Altas Enerx\'ias (IGFAE)}\\
{\small \sl E-15782 Santiago de Compostela, Spain} 

\vspace{2mm}
\vskip 0.2cm
${}^2${\small \sl Centro de F\'isica do Porto} \\
{\small \sl and}\\
{\small \sl Departamento de F\'isica e Astronomia} \\
{\small \sl Faculdade de Ci\^encias da Universidade do Porto}\\
{\small \sl Rua do Campo Alegre 687, 4169-007 Porto, Portugal} 

\end{center}

\vspace{8mm}

\setcounter{footnote}{0}
\renewcommand\thefootnote{\mbox{\arabic{footnote}}}

\begin{abstract}
\noindent
We study the magnetic catalysis of chiral symmetry breaking in the ABJM Chern-Simons matter theory with unquenched flavors in the Veneziano limit. 
We consider a magnetized D6-brane probe in the background of a flavored black hole which includes the backreaction of massless smeared flavors in the ABJM geometry. 
We find a holographic realization for the running of the quark mass due to the dynamical flavors.  We compute several thermodynamic quantities of the brane probe and analyze the effects of the dynamical quarks on the fundamental condensate and on the phase diagram of the model.  The dynamical flavors have an interesting effect on the magnetic catalysis. At  zero temperature and fixed magnetic field, the magnetic catalysis is suppressed for small bare quark masses whereas it is enhanced for large values of the mass. When the temperature is non-zero there is a critical magnetic field, above which the magnetic catalysis takes place. This critical magnetic field decreases with the number of flavors, which we interpret as an enhancement of the catalysis.

\end{abstract}

\newpage

\section{Introduction}\label{intro}

The dynamics of gauge theories in external electromagnetic fields has revealed a rich structure of new phenomena (see \cite{Kharzeev:2013jha} for a recent review). 
One of these effects is the spontaneous symmetry breaking of chiral symmetry induced by a magnetic field, which is 
known as magnetic catalysis \cite{Klimenko,Gusynin:1994re,Gusynin:1994xp,Miransky:2002rp}. It can be understood as due to the fermionic pairing and the effective dimensional reduction which take place in the Landau levels. In strongly interacting systems the holographic duality \cite{Maldacena:1997re} can be used to study this phenomenon \cite{Filev:2007gb} (see \cite{Filev:2010pm,Bergman:2012na} for reviews and further references). The general objective of these holographic studies is to uncover new physical effects of universal nature that are difficult to discover by using more conventional approaches.

In the holographic approach, the matter fields transforming in the fundamental representation of the gauge group are introduced by adding flavor D-branes to the gravity dual. If these flavor branes are treated as probes and their backreaction on the geometry is neglected,  we are in the so-called quenched approximation, which corresponds to discarding quark loops on the field theory side. The magnetic field needed for the catalysis is introduced as a worldvolume gauge field on the D-brane.  From the study of the embeddings of the probe one can extract the $\bar q q$ 
condensate as a function of the quark mass and verify the 
breaking of chiral symmetry induced by the magnetic field.  

To go beyond the probe approximation and to study the effects of quark loops in the holographic approach one has to construct new supergravity duals 
which include the backreaction of the flavor brane sources on the geometry. Finding these unquenched backgrounds is a very difficult problem which can be 
simplified by considering a continuous distribution of flavor branes (see \cite{Nunez:2010sf} for a review of this smearing technique). 
In \cite{Filev:2011mt,Erdmenger:2011bw} the magnetic catalysis for the D3-D7 system with unquenched smeared flavor branes was studied and the effects of 
dynamical flavors on the magnetic catalysis were analyzed.

In this paper we address the problem of the magnetic catalysis with unquenched flavors in the ABJM theory \cite{Aharony:2008ug}. 
The unflavored version of the ABJM model is a  $(2+1)$-dimensional Chern-Simons matter theory with ${\cal N}=6$ supersymmetry, 
whose gauge group is $U(N)\times U(N)$, with Chern-Simons 
levels $k$ and $-k$. It also contains bifundamental matter fields. When the two parameters $N$ and $k$ are large, the ABJM theory can be holographically 
described by the ten-dimensional geometry $AdS_4\times {\mathbb{CP}}^3$ with fluxes. One can naturally add flavor D6-branes extended along the $AdS_4$ 
and wrapping an ${\mathbb{RP}}^3$ submanifold of the internal ${\mathbb{CP}}^3$ \cite{Hohenegger:2009as,Gaiotto:2009tk}. The smeared unquenched background for a large 
number $N_f$ of massless flavors has been constructed in \cite{Conde:2011sw}. These results were generalized in \cite{Jokela:2012dw} to non-zero 
temperature and in \cite{Bea:2013jxa} to massive flavors.  The main advantage of the ABJM case as compared to other 
holographic setups is that the corresponding flavored backgrounds have a good UV behavior without the pathologies present in other unquenched 
backgrounds (such as, for example, the Landau pole singularity of the D3-D7 case). Moreover, in the case of massless flavors the geometry is known 
analytically and is of the form $AdS_{BH_4}\times {\cal M}_6$, where $AdS_{BH_4}$ is a black hole in $AdS_4$ and ${\cal M}_6$ is a squashed version of ${\mathbb{CP}}^3$. This simplicity will allow us to obtain a holographic realization of the Callan-Symanzik equation for the running of the quark mass due to the anomalous dimension generated by the unquenched flavors.

We will carry out our analysis by considering a magnetized D6-brane probe in the geometry \cite{Conde:2011sw,Jokela:2012dw} dual to the ABJM theory 
with unquenched {\it massless} flavors ({\it i.e.}, dynamical {\it sea } quarks), corresponding to the backreaction of a large number $N_f$ of flavor 
D6-branes with no magnetic field.  We are thus neglecting the influence of the magnetic field on the sea quarks. To take this effect into account we 
would have to find the backreaction to magnetized flavor D6-branes, which is an involved problem beyond the scope of this work. 
We will study the system both at zero and non-zero temperature. In both cases we will be able to study the influence of the dynamical 
sea quarks at fully non-linear order in $N_f$.

The rest of this paper  is organized as follows. In Section \ref{The_model} we introduce our holographic model. 
We review the background of \cite{Conde:2011sw,Jokela:2012dw},  study the action of the probe in several coordinate systems and establish the dictionary to relate the holographic parameters to the physical mass and condensate. 
In Section \ref{The_thermo} we obtain  the different thermodynamic properties of the magnetized brane and we find analytic results in some particular limiting cases. 
Section \ref{sec:catalysis} is devoted to the study of the phase diagram and of the magnetic catalysis of chiral symmetry breaking. 
Finally, in Section \ref{conclusion} we summarize our results and discuss some possible research directions for the future. Appendix \ref{appendix} contains
the derivation of the holographic dictionary for the condensate at zero temperature.

\section{Holographic model}
\label{The_model}

In this section  we will recall the background of type IIA supergravity dual to unquenched massless flavors in the ABJM Chern-Simons matter theory at non-zero temperature. 
This background was obtained \cite{Conde:2011sw,Jokela:2012dw} by including the backreaction of $N_f$ flavor D6-branes, which are continuously distributed in the internal 
space in such a way that the system preserves  ${\cal N}=1$ supersymmetry at zero temperature. This smearing procedure is a holographic implementation of the so-called 
Veneziano limit \cite{Veneziano:1976wm}, in which both $N$ and $N_f$ are large. 
As the smeared flavor branes are not coincident the flavor symmetry is $U(1)^{N_f}$ rather than $U(N_f)$. 

To study magnetic catalysis in this gravity dual with unquenched flavors, we will add an additional flavor D6-brane probe with a magnetic field in 
its worldvolume. We will obtain the action of this probe and introduce various systems of coordinates which are convenient to describe the embeddings of the 
brane, both at zero and non-zero temperature.

\subsection{Background metric}
Our model consists of a probe D6-brane in the smeared flavored ABJM background of \cite{Conde:2011sw,Jokela:2012dw}, oriented such that their intersection is $(2+1)$-dimensional.  
We will begin by laying out our conventions and reviewing the background geometry.
The metric of the background is \cite{Conde:2011sw,Jokela:2012dw}
\be\label{eq:10dmetric} 
 ds_{10}^2 =  L^2 \left(-h r^2 dt^2+r^2(dx^2+dy^2)+\frac{dr^2}{hr^2}\right)+\frac{L^2}{b^2}\left(q ds^2_{\mathbb{S}^4}+(E^1)^2+(E^2)^2\right) \ ,
\ee 
where $L$ is a constant radius and the blackening factor is   $h(r)=1-\frac{r_h^3}{r^3}$, with $r_h$ constant.  In our conventions all coordinates are dimensionless and $L$ has dimension of length.  The Bekenstein-Hawking temperature  $T$ of the black hole is related to $r_h$ as $T=\frac{3 r_h}{4\pi}$. Notice that  $T$ is dimensionless (the physical temperature is   $T/\sqrt{\alpha'}$). 
The internal metric in (\ref{eq:10dmetric}) is a deformation of the Fubini-Study metric of ${\mathbb C}{\mathbb P}^3$, represented as an 
${\mathbb S}^2$-bundle over ${\mathbb S}^4$. This deformation is generated by the backreaction of the massless flavors and introduces a  relative squashing $q$ between the ${\mathbb S}^2$ fiber, corresponding to the two one-forms $E^1$ and $E^2$, and the ${\mathbb S}^4$ base. We write the metric on the four-sphere in (\ref{eq:10dmetric}) as
\be
ds^2_{\mathbb{S}^4} = \frac{4}{(1+\xi^2)^2}\,
\Big[d\xi^2+\xi^2\,\sum_{i=1}^3 ( \omega^i)^2\,\Big]\,\,,
\ee
where $0\le \xi<\infty$ is a non-compact coordinate and the $\omega^i$ are 
$SU(2)$ left-invariant one-forms satisfying $d\omega^i=\frac{1}{2}\epsilon_{ijk}\,\omega^j\wedge \omega^k$. The  ${\mathbb S}^2$  will be represented by the ordinary polar coordinates $0\le \theta<\pi$ and $0\le\varphi <2\pi$, in terms of which 
$E^1$ and $E^2$ can be written as
\bea
 E^1 & = & d\theta+\frac{\xi^2}{1+\xi^2}\left(\sin\varphi\omega^1 -\cos\varphi\omega^2\right) \\
 E^2 & = & \sin\theta\left(d\varphi-\frac{\xi^2}{1+\xi^2}\omega^3 \right)+\frac{\xi^2}{1+\xi^2}\cos\theta\left( \cos\varphi\omega^1+\sin\varphi\omega^2\right) \ .
\eea

The metric (\ref{eq:10dmetric}) has two parameters $q$ and $b$ which deserve pronunciation. The parameter $q$ is a constant squashing factor
of the internal $\mathbb{CP}^3$ sub-manifold, whereas $b$ represents the relative squashing between the internal space and the $AdS_{BH_4}$ part of the metric.
The explicit expressions for the factors $q$ and $b$ of the smeared solution of \cite{Conde:2011sw,Jokela:2012dw} are:
\bea
 q & = & 3+\frac{3}{2}\hat\epsilon -2\sqrt{1+\hat\epsilon+\frac{9}{16}\hat\epsilon^2}   \\
 b & = & \frac{2q}{q+1}\ ,\label{b}
\eea
where $\hat\epsilon$ is  the flavor deformation parameter, which depends on the number of flavors $N_f$ and colors $N$, as well as the 't Hooft coupling $\lambda=N/k$, via
\be
 \hat \epsilon = \frac{3N_f}{4k} = \frac{3}{4}\frac{N_f}{N}\lambda \ .
 \label{hat-epsilon_def}
\ee
The radius $L$ in (\ref{eq:10dmetric}) is also modified by the backreaction of the flavors. Indeed, it can be written as \cite{Conde:2011sw}
\be
L^2\,=\,\pi\,\sqrt{2\lambda}\,\,\sigma\,\alpha'
\,\,,
\label{flavored-AdS-radius}
\ee
where $\sigma$ is the  so-called screening function, which determines the correction of the radius with respect to the unflavored case and is given by the following function of the deformation parameter
\be
\sigma\,\equiv\,\sqrt{
\frac{2-q}{q\big[q+(1+\hat\epsilon)(q-1)\big]}}\,\,b^2\,=\,
\frac{1}{4}\frac{q^{\frac{3}{2}}\,\,(2-q)^{\frac{1}{2}}\,\,(1+\hat\epsilon+q)^2}{\big[q+(1+\hat\epsilon)(q-1)\big]^{\frac{5}{2}}}\,\,.
\label{screening-sigma}
\ee
We note that most of the equations that we will manipulate in this paper only depend on $b$, in which case it suffices to keep in mind that $b$ is monotonously increasing between $1$ and $5/4$ as one dials $\hat\epsilon = 0$ to $\infty$. Moreover, $\sigma=1$ 
for  $\hat\epsilon = 0$ and it vanishes as $1/\sqrt{\hat \epsilon}$ when $\hat\epsilon$ is large. 

The type IIA supergravity solution of \cite{Conde:2011sw,Jokela:2012dw} also contains a constant dilaton $\phi$, given by
\be
e^{-\phi} = \frac{b}{4}\frac{1+\hat\epsilon+q}{2-q}\,\frac{k}{L}\,\sqrt{\alpha'}
\,\,,
\label{dilaton-flavored-squashings}
\ee
as well as RR forms $F_2$ and $F_4$. In this paper we will only need the seven-form potential $C_7$ of $F_8=-*F_2$. To avoid unnecessary notation, 
we shall only present its pullback in the subsection to follow.

\subsection{D6-brane action}

Next we will add a probe D6-brane in this background, extended along the Minkowski  and radial coordinates and 
wrapping a three cycle  ${\cal C}_3 \simeq{\mathbb{RP}}^3$ inside the internal manifold. 
The cycle  ${\cal C}_3$ extends along two directions of the ${\mathbb S}^4$ and one direction of the ${\mathbb S}^2$ fiber. 
It can be characterized by requiring that the pullbacks of two of the one-forms $\omega^i$ vanish (say, $\omega^1$ and $\omega^2$) 
and that the angle $\theta$ of the  ${\mathbb S}^2$ is a function of the radial variable. By a suitable choice of 
coordinates,\footnote{Let us require the pullbacks $\hat\omega^1=\hat\omega^2=0$ and parameterize  $\hat\omega^3=d\hat\psi$. 
Then, $\alpha$, $\beta$, and $\psi$ are defined as: $\xi  =:  \tan\left(\frac{\alpha}{2}\right)$,  $\beta := \frac{\hat\psi}{2}$,  and $\psi := \varphi-\frac{\hat\psi}{2}$. 
} the induced metric on the D6-brane worldvolume can be written as
\bea
&&d\hat s^2_7\,=\,
-L^2r^2\,dt^2+
L^2\,r^2\big[\,(dx^1)^2+(dx^2)^2\,\big]\,\,+\,
\frac{L^2}{r^2}\Big[\,1+ \frac{r^2}{b^2}\,\dot\theta^2\,\Big]\,
dr^2+\rc\rc
&&\qquad\qquad\qquad
+\frac{L^2}{b^2}\,\Big[\, q d\alpha^2\,+\,q\,\sin^2\alpha\, d\beta^2\,+\,
\sin^2\theta\,
\big(\,d\psi\,+\,\cos\alpha\,d\beta\,\big)^2\,\Big]
\,\,,
\label{induced-general}
\eea
where $\dot\theta=d\theta/dr$ and $0\le\alpha <\pi$, $0\le\beta,\psi<2\pi$.

The D6-brane action has two contributions. As usual there is the Dirac-Born-Infeld term, but we also have a Chern-Simons term due to the pullback of  the RR seven-form potential 
\be
S_{D6}\,=\,-T_{D6}\,e^{-\phi}\,\int d^{7}\zeta \sqrt{-\det(g_7+F)}\,+\,
T_{D6}\,\int \hat C_7\,\,,
\eeq
where the $\zeta$'s are the coordinates of the induced metric and $F=dA$ is the strength of the worldvolume gauge field. The explicit form of the the pullback of 
$C_7$ is\cite{Jokela:2012dw} 
\be
 \hat C_7 = \frac{L^7q}{b^3}e^{-\phi}d^3x\wedge\left[\frac{hr^3}{b}\sin\theta\cos\theta \dot\theta+r^2\sin^2\theta+L_2(r)  \right]\wedge dr \wedge 
  \Xi_3 \ ,
\ee
where $\Xi_3  =  \sin\alpha\, d\alpha\wedge d\beta\wedge d\psi$  and $\int dr L_2(r) = \frac{r_h^3}{4b}$.  To write $C_7$ we have chosen a particular 
gauge which leads to a finite renormalized action with consistent thermodynamics. 
In this paper we will consider a background magnetic field described by a spatial component of the D6-brane gauge field:
\beq
A_{x^{2}}\,=\,x^1\,L^2\,B \ .
\eeq
In our conventions, the quantity $B$, as well as $x^1$, is dimensionless. Notice also  that the physical magnetic field is related to $B$ as
\be
B_{phys}={L^2\,B\over \alpha'^2}\,=\,{\pi\sqrt{2\lambda}\,\sigma B\over \alpha'}\,\,.
\label{Bphys_def}
\ee

A straightforward computation for the full action yields:
\be\label{eq:action}
 S = -{\cal N}\int d^3x \left\{\frac{4b}{r_h^3}\int dr r^2\sin\theta\left(\sqrt{1+\frac{B^2}{r^4}}\sqrt{1+h\left(\frac{r}{b}\right)^2\dot\theta^2}-\sin\theta-h\frac{r}{b}\cos\theta\dot\theta\right)-1 \right\}\ ,
\ee
where  the prefactor is
\be
 {\cal N}  =  \frac{2\pi^2 \,r_h^3\, L^7\,q}{b^4}\,T_{D6}\,e^{-\phi} =  \,\frac{2\sqrt 2\pi^2(2-b)\,b\,\sigma}{27}\, N\,\sqrt\lambda \,\,T^3\,\,.
 \ee
For later use we also define:
\be
 {\cal N}_r  =  \frac{4b}{r_h^3}\,{\cal N} = \frac{(2-b)\,b^2\,\sigma}{4\sqrt 2\pi}\,\,
  \frac{N^{3/2}}{\sqrt k} \ .
\ee
The equation of motion for the embedding scalar is thus,
\be
 \partial_r\left(g \left(\frac{r}{b}\right)^2 \left(1+\frac{B^2}{r^4}\right)\dot\theta \right) = r^2\,\left(\frac{3}{2b}-1+\frac{hr^2}{2g}\right)\sin 2\theta \ ,
\ee
where we have defined
\be
 g = \frac{hr^2\sin\theta}{\sqrt{1+\frac{B^2}{r^4}}\sqrt{1+h\left(\frac{r}{b}\right)^2\dot\theta^2}} \ .
\ee

The above equation of motion has generically two kinds of solutions. The first kind are embeddings that penetrate the black hole horizon, those 
we shall call black hole (BH) embeddings. The
other kind are Minkowski (MN) embeddings, which terminate smoothly above the horizon at some $r_0>r_h$. Examples of both kind are the following. Clearly, 
the equation of motion is satisfied with trivial constant angle BH embeddings $\theta=0,\pi/2$. The equation of motion possesses a supersymmetric
MN solution $\cos\theta(r)=\left(\frac{r_0}{r}\right)^b$ at $r_h=0$ and $B=0$. Away from zero temperature and vanishing magnetic field, this solution
has to be analyzed numerically. Our focus in this article is to study how these two types of solutions map out the phase space as both the $T$ and $B$ 
are dialed, and the interesting effects from the variation of the number of background flavors (essentially $b$). Before we will get absorbed in
analyzing several aspects of the system, we wish to introduce new parameterizations better suited for the analyses.

\subsection{Parameterization at non-zero temperature}

It is useful to introduce another parameterization as discussed in \cite{Jokela:2012dw}. Let us introduce a system with isotropic Cartesian-like coordinates
\bea
    R & = & u\cos\theta \label{eq:cartR}\\
 \rho & = & u\sin\theta \label{eq:cartrho}  \ ,
\eea
where the new radial coordinate $u$ is related to the old one as
\be
 u^{\frac{3}{2b}} = \left(\frac{r}{r_h}\right)^{\frac{3}{2}}+\sqrt{\left(\frac{r}{r_h}\right)^3-1} \ .
\ee
We also define the functions $f$ and $\tilde f$ as
\bea
 f & = & 1-u^{-3/b} 
 \label{f_def}
 \\
 \tilde f & = & 1+u^{-3/b} \ .
 \label{ftilde_def}
\eea
We also rescale the magnetic field as follows:
\be\label{eq:Bhat}
 \hat B = 2^{4/3}\frac{B}{r_h^2} \ .
\ee
After these mappings the action becomes
\bea
 S & = & -{\cal N}\int d^3x \Bigg\{\int d\rho \rho f\tilde f u^{3/b-2}\Bigg(\sqrt{1+\frac{\hat B^2}{\tilde f^{8/3}u^{4/b}}}\sqrt{1+R'^2}-1 \nonumber\\
  & & \qquad\qquad\qquad +\left(\frac{f}{\tilde f}-1\right)\frac{R}{u^2}\left(\rho R'-1\right)  \Bigg)-1 \Bigg\}\ ,\label{eq:Taction}
\eea
where it is understood that $u=\sqrt{\rho^2+R^2}$.

A generic solution to the equation of motion following from the action (\ref{eq:Taction}), behaves close to the boundary as:
\be
 R = m + \frac{c}{\rho^{3/b-2}} + \ldots \ , \ \rho\to\infty \ ,
\ee
where $m$ is related to the quark mass and $c$ is proportional to the vacuum expectation value $\langle \bar\psi\psi\rangle$ (see below).

\subsection{Parameterization at zero temperature}
\label{zeroT_section}

At zero temperature we also make use of the Cartesian-like coordinates as in (\ref{eq:cartR}) and (\ref{eq:cartrho}), but with
\be
 u = r^b \ .
\label{u-r}
\ee
The action (\ref{eq:action}) maps to
\be
 S = -\frac{{\cal N}_r}{b} \int d\rho \rho u^{3/b-2}\left\{\sqrt{1+\frac{B^2}{u^{4/b}}}\sqrt{1+R'^2}-1  \right\} \ ,
\ee
where it is understood that $u=\sqrt{\rho^2+R^2}$. 
We can scale out the $B$ as follows:
\bea
 u^{4/b} & = & B^2\tilde u^{4/b} \to u = B^{b/2}\tilde u \\
 R & = & B^{b/2}\tilde R \\
 \rho & = & B^{b/2}\tilde \rho \ .
\eea
This leads us to
\be\label{eq:zeroTaction}
 S = -\frac{B^{3/2}}{b}{\cal N}_r\int d\tilde\rho \tilde\rho \tilde u^{3/b-2}\left\{ \sqrt{1+\frac{1}{\tilde u^{4/b}}}\sqrt{1+\tilde R'^2}-1  \right\}
\ee
and to the following asymptotic behavior of the embedding function
\be\label{eq:zeroTasymp}
 R\sim m_0+\frac{c_0}{\rho^{3/b-2}} \to \tilde R\sim \tilde m_0 + \frac{\tilde c_0}{\tilde\rho^{3/b-2}} \ ,
\ee
where we defined the dimensionless quantities $\tilde m_0$ and $ \tilde c_0$ as:
\bea
 \tilde m_0 & \equiv & B^{-b/2} m_0\label{tilde_m_0} 
 \label{tilde_m_0_def}\\
 \tilde c_0 & \equiv & B^{(b-3)/2}c_0\label{tilde_c_0}\ .
  \label{tilde_c_0_def}
\eea
The parameters $m_0$ and $c_0$ can be related to the quark mass $m_q$ and the quark condensate at zero temperature (denoted   by $\left\langle {\cal O}_q\right\rangle_0$). The corresponding relation is worked out in the next section and in appendix 
\ref{appendix}. 

\subsection{Running mass and condensate}
\label{runnning_mass_section}

The asymptotic value of the embedding function $R$ should be  related to the quark mass. 
To find the precise relation we will consider a fundamental string stretched in the $R$ direction and ending on the flavor brane. 
The quark mass is just the Nambu-Goto action of the string per unit time. 
While carrying out this computation we should take into account that we are dealing 
with a theory with unquenched quarks in which the quark mass $m_q$ acquires an anomalous dimension $\gamma_m$ and $m_q$ therefore 
runs with the scale according to the corresponding Callan-Symanzik equation. 
In our holographic setup the value of  $\gamma_m$ was found in \cite{Conde:2011sw,Jokela:2012dw} and is simply related to the squashing parameter $b$:
\beq
\gamma_{m}=b-1\,\,.
\label{gamma_m}
\eeq
In order to find the scale dependence of $m_q$, we consider a fundamental string located at the point $\rho=\rho_*$. We will start by considering the zero temperature case. Notice that $\rho$ is the holographic coordinate in our setup and, therefore, it is natural to think that the value of $\rho_*$ determines the energy scale. The induced metric on a string worldsheet extended in  $(t,R)$ at $\rho=\rho_*$  when $T=0$ is given by:
\beq
ds^2_{2}\,=\,-L^2\,\big[R^2+\rho_*^2\big]^{{1\over b}}\,dt^2\,+\,
{L^2\over b^2}\,{dR^2\over R^2+\rho_*^2}\,\,.
\eeq
The running quark mass at zero temperature is  then defined as:
\bear
m_q & = & {1\over 2\pi (\alpha')^{{3\over 2}}}\,
\int_0^{m_0}\,\sqrt{-\det g_2}\,\,dR\,=\,
\sqrt{{\lambda\over 2}}\,\,{\sigma\over b\sqrt{\alpha'}}\,
\int_0^{m_0}
\big[R^2+\rho_*^2\big]^{{1\over 2b}-{1\over 2}}\,dR \nonumber\\
& = & \sqrt{{\lambda\over 2}}\,\,{\sigma\over b\sqrt{\alpha'}}\,
\, m_0\,\,\rho_*^{{1\over b}-1}\,
{}_{2}F_{1}\Big(\,{1\over 2}, {\gamma_m\over 2 b};{3\over 2};
-{ m_0^2\over \rho_*^{2}}\,\Big)\,\,.
\label{mq_integrated}
\eear
Notice that, in the unflavored case $b=1$, $\gamma_m=0$ and the effective mass $m_q$ is independent of the scale parameter $\rho_*$, as it should. 
To determine the precise relation between $\rho_*$ and the energy scale $\Lambda$, let us consider the relation (\ref{u-r}) between the 
coordinate $u$ and the canonical $AdS_4$  radial coordinate $r$. 
Taking into account that $u\approx \rho$ in the UV, it is natural to identify $r$ with the energy scale and define $\Lambda$ as:
\beq
\Lambda\equiv \rho_*^{{1\over b}}\,\,.
\label{Lambda-def}
\eeq
The dependence of $m_q$ on $\Lambda$ can be straightforwardly inferred from (\ref{mq_integrated}). 
Moreover, from the integral representation in (\ref{mq_integrated}) we can readily obtain an evolution equation for $m_q$
\beq
{\partial\, m_q\over \partial \log\Lambda}\,=\,m_q\,-\,{\sigma\over \sqrt{ \alpha'}}\,
\sqrt{{\lambda\over 2}}\,
{m_0\over 
(\Lambda^{2b}+m_0^2)^{{\gamma_m\over 2b}}}\,\,.
\label{evo_eq_mq}
\eeq
Clearly, the second term in (\ref{evo_eq_mq}) incorporates  the flavor effects on the running of $m_q$.  
In the UV regime of large $\Lambda$ we can just neglect $m_0^2$ in  the denominator of (\ref{evo_eq_mq}). 
The solution of this UV equation can be obtained directly or by taking the large $\Lambda$ limit of (\ref{mq_integrated}). We get
\beq
{m_q\,\sqrt{\alpha'}\over \sqrt{\lambda}}\approx {\sigma\over \sqrt{2}\,\,b}\,\,m_0\,\,
\Lambda^{-\gamma_m}\,\,,
\label{UV_running_mass}
\eeq
which shows that in the UV $m_q$ and $m_0$ are proportional and that  the running of $m_q$ with  the scale $\Lambda$ is controlled by the mass  anomalous dimension $\gamma_m$. Notice that the UV mass  (\ref{UV_running_mass}) satisfies:
\beq
{\partial\, m_q\over \partial \log\Lambda}\,=\,-\gamma_m\,\,m_q\,\,,
\label{CS-eq}
\eeq
which is just the Callan-Symanzik equation for the effective mass.  

The analysis carried out above for $m_q$ is independent of the value of the magnetic field $B$. When $B\not=0$ it is convenient to write the solution of the evolution equation in terms of the reduced mass parameter $\tilde m_0$ defined in (\ref{tilde_m_0_def}). We get:
\beq
{m_q\,\sqrt{\alpha'}\over \sqrt{\lambda}}=
 {\sigma\over \sqrt{2}\,\,b}\,\,B^{{b\over 2}}\,\tilde m_0\,\,
\Lambda^{-\gamma_m}\,
{}_{2}F_{1}\Big(\,{1\over 2}, {\gamma_m\over 2 b};{3\over 2};
-{B^b\,\tilde m_0^2\over \Lambda^{2(1+\gamma_m)}}\,\Big)\,\,.
\label{m_q_tilde_m_0}
\eeq

To find the relation between the parameter $c_0$ in (\ref{eq:zeroTasymp})  and the condensate we have to compute the derivative of the free energy with respect to the bare quark mass $\mu_q^0$, which is the  quark mass without the screening effects due to the quark loops. These effects are encoded in the functions $\sigma$ and $b$. By putting $\sigma=b=1$, which corresponds to taking $\hat \epsilon=0$, we switch off the dressing due to the dynamical flavors. Accordingly, to get  $\mu_q^0$ in terms of $\tilde m_0$ we just take $\sigma=b=1$ on the right-hand side of (\ref{m_q_tilde_m_0}). We get 
\beq
\mu_q^0\,=\,\sqrt{{\lambda\over 2}}\,{\sqrt{B}\,\tilde m_0\over \sqrt{\alpha'}}\,\,.
\label{mu_q_zeroT}
\eeq
Notice that the value of $\tilde m_0$ does not depend on the magnetic field, which is factorized in the action (\ref{eq:zeroTaction}). Therefore, the dependence of $\mu_q^0\sim \sqrt{B} $ on the field  $B$ is the same as in the unflavored case, as it should.

The explicit calculation of the vacuum expectation value $\left\langle {\cal O}_q\right\rangle_0$ has been performed in Appendix \ref{appendix}, with the result:
\be
-\frac{\left\langle {\cal O}_q\right\rangle_0\,\alpha'}{N}  =  
\frac{(3-2b)(2-b)}{4\pi }\,\sigma \,
 B^{\frac{\gamma_m}{2}}\,
 c_0 = \frac{(3-2b)(2-b)}{4\pi }\,\sigma\, B\,\tilde c_0 \,\,.
 \label{zeroTdictc} 
 \ee
Eqs. (\ref{mu_q_zeroT})  and (\ref{zeroTdictc}) constitute the basic  dictionary in our analysis of the chiral symmetry  breaking at zero temperature. 

For non-zero temperature we shall  proceed as in the $T=0$ case. The induced metric for the fundamental string  extended in $R$ at $\rho=\rho_*$  is now
\beq
ds^2_{2}\,=\,-{L^2\,r_h^2\over 2^{{4\over 3}}}\,
\big[R^2+\rho_*^2\big]^{{1\over b}}\,\big[f_*(R)\big]^{2}\,\big[\tilde f_*(R)\big]^{-{2\over 3}}
dt^2\,+\,
{L^2\over b^2}\,{dR^2\over R^2+\rho_*^2}\,\,,
\eeq
where $f_*(R)$ and $\tilde f_*(R)$ are the functions defined in (\ref{f_def}) and (\ref{ftilde_def}) at $\rho=\rho_*$. Accordingly, 
the running quark mass at $T\not =0$ is now given by  an integral extended 
from the horizon (for $\rho_*<1$) to $R=m$:
\beq
m_q^{T}\,=
\sqrt{{\lambda\over 2}}\,\,{\sigma\over b\sqrt{\alpha'}}\,{r_h\over 2^{{2\over 3}}}
\int_{R_h}^{m}
\big[R^2+\rho_*^2\big]^{{1\over 2b}-{1\over 2}}\,
f_*(R)\,\big[\tilde f_*(R)\big]^{-{1\over 3}}\,
dR\,\,,
\eeq
where $R_h\,=\,\sqrt{1-\rho_*^2}$ for $\rho_*<1$ and $R_h=0$ otherwise. 
We have not been able to integrate analytically this expression for arbitrary values of 
$\rho_*$.  In order to relate $\rho_*$ with the scale $\Lambda$ we recall that, in the UV, $\rho^{{1\over b}}\approx u^{{1\over b}}\approx 2^{{2\over 3}}\,r/r_h$. 
Therefore, identifying again $r$ with $\Lambda$, we have:
\beq
\Lambda\,=\,2^{-{2\over 3}}\,r_h\,\rho_*^{{1\over b}}\,\,.
\label{Lambda-def-finiteT}
\eeq
We readily obtain in the UV domain ($\Lambda\gg 1$): 
\beq
{m_q^{T}\,\sqrt{\alpha'}\over \sqrt{\lambda}}\approx {\sigma\over \sqrt{2}\,\,b}\,
\,{r_h^b\over 2^{{2b\over 3}}}\,
\,m\,\,
\Lambda^{-\gamma_m} \ .
\label{UV_running_mass_T}
\eeq
This UV function $m_q^{T}$ also satisfies the Callan-Symanzik equation (\ref{CS-eq}). Actually, it is easy to relate in the UV the effective mass $m_q^{T}$ to its zero temperature counterpart. In order to establish this connection,  let us connect  
$m_0$ and $c_0$ with the zero temperature limit of $m$ and $c$. To find these relations we recall that these parameters characterize the leading and subleading UV behaviors of the embedding function. 
From this observation it is easy to prove that
\beq
m^{\frac{1}{b}}\approx 2^{\frac{2}{3}}\,r_h^{-1}\,m_0^{\frac{1}{b}}\,\,,
\qquad\qquad
c\,\approx \,2^{2-\frac{2b}{3}}\,r_h^{b-3}\,c_0
\,\,,
\qquad\qquad
(T\to 0)\,\,.
\label{mc-mc-zero}
\eeq
By using the relation  between $m$ and $m_0$ written in (\ref{mc-mc-zero}), we see that  $m_q$ is just the limit of   $m_q^{T}$ as $T\to 0$:
\beq
m_q\,=\,\lim_{T\to 0}\,m_q^{T}\,\,.
\eeq

The bare mass at non-zero temperature  $\mu_q$ is obtained from the unflavored limit of  the UV running mass (\ref{UV_running_mass_T}). We get \cite{Jokela:2012dw}:
\beq
\mu_q\,\sqrt{\alpha'}\,=\,{2^{{1\over 3}}\,\pi\over 3}\,\sqrt{2\lambda}\,T\,m\,\,.
\eeq
The resulting vacuum expectation value at $T\not=0$ has been obtained in appendix D of \cite{Jokela:2012dw} and is 
given by:
\beq
 -\frac{\left\langle {\cal O}_q\right\rangle\,\alpha'}{N}  =  \frac{2^{2/3}\pi(3-2b)(2-b)}{9}\,\sigma\,T^2\, c \ .\label{Tdictc}
\eeq
The relation between the condensates at non-zero and zero temperature is similar to the one corresponding to the masses. 
Indeed,  by using (\ref{mc-mc-zero}) we get that $\left\langle {\cal O}_q\right\rangle_0$  is given by the following zero temperature limit:
\beq\label{eq:dictrelation}
\left\langle {\cal O}_q\right\rangle_0\,=\,
\lim_{T\to 0}\Big[ \hat B^{\frac{\gamma_m}{2}}\,
\left\langle {\cal O}_q\right\rangle\Big]\,\,.
\eeq

The relation (\ref{eq:dictrelation}) is very natural from the point of view of the renormalization group. Indeed, $\langle {\cal{O}}_q\rangle$ and 
$\langle {\cal{O}}_q\rangle_0$ are dimensionful quantities defined at scales determined by the temperature and the magnetic field, respectively. The quotient
$\langle {\cal{O}}_q\rangle/\langle {\cal{O}}_q\rangle_0$ should be given by the ratio of these two energy scales (which is basically $\sqrt{\hat B}$) raised to some power which,
following the renormalization group logic, should be the mass anomalous dimension, as in (\ref{eq:dictrelation}).

In the rest of this paper, we will use units in which $\alpha'=1$. The appropriate power of $\alpha'$ can be easily obtained in all expressions by looking at their units. 


\section{Some properties of the dual matter}
\label{The_thermo}

We will discuss many of the characteristics of the dual matter as described by the gravitational system. The BH phase describes
typical metallic behavior. The phase is nongapped to charged and neutral excitations. For example, the former can be easily verified
by the standard DC conductivity calculation \cite{Karch:2007pd} in a simple generalization of our model by introducing a non-vanishing charge density
on the probe. The MN phase, on the other hand, behaves like an insulator: it is gapped to both neutral and charged excitations; 
the latter can be checked by the conductivity calculation of \cite{Bergman:2010gm} and the former by fluctuation analysis.
The interplay between these two phases in the presence of a charge density makes an interesting story, which will be addressed in a future work.

In the absence of the magnetic field, the thermodynamic properties of the system were discussed in great detail in \cite{Jokela:2012dw}. Here we are more 
interested in the magnetic properties and on the effects that the magnetic field will bear. We will break this narrative in two parts, so that
in this section we will constrain ourselves to the case where we have analytic control and in the next section we will confront
the numerical side of the story, most relevantly the magnetic catalysis.

\subsection{Thermodynamic functions}
The free energy of the system is obtained from evaluating the 
Wick rotated on-shell action (\ref{eq:action}). 
As discussed in \cite{Conde:2011sw,Jokela:2012dw},
the free energy is finite albeit subtle at non-zero temperature; there is no need to invoke holographic renormalization to get rid off infinities.
The free energy of the probe is identified with the Euclidean on-shell action $S_E$, through the relation $F=T\,S_E$. In the calculation of $S_E$
we integrate over both the Euclidean time and the non-compact two-dimensional space. Since the latter integration gives rise to an (infinite) two-dimensional volume $V_2$, 
from now on we divide all the extensive thermodynamic quantities by $V_2$ and deal with densities.
The free energy  density $F$ can be written as
\begin{equation}
\frac{F}{{\cal N}}\,=\,{\cal G}(m,  \hat{B})\,-\,1 \, .
\label{F-calG}
\end{equation}
The explicit expression for the function ${\cal G}(m, \hat B)$  can be obtained from the action of the D6-brane probe. For MN embeddings it is 
more convenient to use $R(\rho)$ as embedding function. From the expression \eqref{eq:Taction} of the action in these variables, ${\cal G}(m, \hat B)$ is given by
\begin{eqnarray}
{\cal G}(m, \hat{B})\, & = & \,
\int_{0}^{\infty}d\rho\,
\rho\,\big[\,\rho^2\,+\,R^2\,\big]^{\frac{3}{2b}-1}f\,\tilde f\,
\Bigg[\sqrt{1+ R'^2}\,\sqrt{1 \, + \, \frac{\hat{B}^2}{\tilde f^\frac{8}{3}} \,\big[\,\rho^2\,+\,R^2\,\big]^{-\frac{2}{b}} }
\nonumber \\
&- & 
1\,+\,\left(\frac{f}{\tilde f}-1\right)\,\frac{R}{\rho^2+R^2}\,(\rho R'-R)\,\Bigg] \, .
\label{calG}
\end{eqnarray}
For black hole embeddings it is better to use the $\theta=\theta(r)$ parameterization and represent  ${\cal G}(m, \hat B)$ as:
\be
 {\cal G}(m, \hat B)\,=\,\frac{4b}{r_h^3}
 \int_{r_h}^{\infty}dr r^2 \sin^2\theta\Bigg[\sqrt{1+
 \Big(\frac{r_h}{2^{\frac{2}{3}} r}\Big)^{4}\,\hat B^2}\,
 \sqrt{1+h\,\Big(\frac{r}{b}\Big)^2\,\dot\theta^2}-\sin\theta\,-\,{h \frac{r}{b}}\,\cos\theta\,\dot\theta\,\Big]\,\,.
 \label{calG_BH}
 \ee

The fact that the system under study is defined at a fixed temperature and magnetic field implies that the appropriate thermodynamic potential is 
\begin{equation}\label{def-F}
dF \, = \, - \, s\,dT \, - \, {\cal M}\, dB \, ,
\end{equation}
where $s$ is the entropy density and ${\cal M}$ is the magnetization of the system. Following \eqref{def-F}, the entropy density $s$ is given by the following expression
\be\label{def-s}
s  = -\left(\frac{\partial F}{\partial T}\right)_{B} \, = \, -  \, \frac{{\cal N}}{T} \, 
\left[\frac{3 \, F}{\cal N} \,  + \, T\, \left(\frac{\partial {\cal G}}{\partial m}\right)_{\hat{B}} \left(\frac{\partial m}{\partial T}\right)_{B} 
+ \, T\, \left(\frac{\partial {\cal G}}{\partial \hat{B}}\right)_{m} \left(\frac{\partial \hat{B}}{\partial T}\right)_{B} \right]\,\,. 
\ee
Let us compute the different derivatives on the right-hand side of (\ref{def-s}). First of all we use that \cite{Jokela:2012dw}
\be
\frac{\partial {\cal G}}{ \partial m}\,=\,\frac{2b-3}{b}\,c\,\,,
\label{calG-m_derivative}
\ee
and that $T\,\partial m/\partial T=-bm$, as follows from (\ref{UV_running_mass_T}) when $m_q^T$ and $\Lambda$ are fixed.  Moreover, 
we  define the function ${\cal J}(m,\hat B)$ as
\be
{\cal J}(m,\hat B)\,\equiv\,\frac{1}{\hat B}\,
\frac{\partial   {\cal G}(m, \hat B)}{ \partial \hat B}\,\,.
\ee
Then, taking into account the $\hat B\propto T^{-2}$ temperature dependence of the rescaled magnetic field in (\ref{eq:Bhat}), we get:
\be
T\,\frac{s}{{\cal N}} =  3\,- \,3\,{\cal G}(m,\hat B)\,+\,2\,\hat B^2\,{\cal J}(m,\hat B)
\,-\, (3-2b)\,c\,m\,\,.
\label{s_general}
\ee
For Minkowski embeddings, ${\cal J}(m,\hat B)$ is explicitly given by the following integral:
\be
{\cal J}(m,\hat B) \equiv \int_{0}^{\infty} d\rho \big[\,\rho^2\,+\,R^2\,\big]^{\frac{1}{2b}-1}\,f\,\tilde f^{-\frac{1}{3}} \frac{\sqrt{1+R'^2}}{\sqrt{\hat B^2\,+\,\big(\,\rho^2\,+\,R^2\,\big)^\frac{2}{b}\tilde f^{\frac{8}{3}}}} \ ,
\label{calJ_MN}
\ee
whereas for a black hole embedding we have:
\be
{\cal J}(m,\hat B)\,=\,{b\,r_h\over 2^{{2\over 3}}}\,\,
\int_{r_h}^{\infty}dr  \sin^2\theta\,
{\sqrt{1+h\,\Big({r\over b}\Big)^2\,\dot\theta^2}\over 
\sqrt{r^4\,+\,\Big({r_h\over 2^{{2\over 3}} }\Big)^{4}\,\hat B^2}}\,\,.
\label{calJ_BH}
\ee

The internal energy density $E$ can be computed from the relation $E=F+T s$, with the result
\be
{E\over {\cal N}} \,=\,2\, -2\,{\cal G}(m,\hat B)\,+\,2\,\hat B^2\,{\cal J}(m,\hat B)\,-\,
(3-2b)\,c\,m\,\,.
\label{E-calG}
\ee
The heat capacity density $c_v$ is defined as $c_v=\partial E/\partial T$. 
Computing explicitly the derivative  of  $E$ with respect to the temperature in (\ref{E-calG}), and using (\ref{calG-m_derivative}), we arrive at the  following expression:
\bear
&&T\,{c_v\over {\cal N}}\,=\,
2\,T\,{s\over {\cal N}}\,-\,2\,\hat B^2\,\Big({\cal J}(m,\hat B)\,+\,
2\hat B\,{\partial {\cal J}(m,\hat B) \over \partial\hat B}\,\Big)\,+
\rc\rc
&&\qquad\qquad\qquad
+(2b-3)\Bigg[
\Big(\,3-b-b\,{\partial ( \log c)\over \partial (\log m)}\,\Big)\,c\,m\,-\,
4m\,\hat B{\partial c\over \partial\hat B}\Bigg]
\,\,. 
\label{cv-s-general}
\eear

In order to holographically investigate the joint effect of the presence of flavors and magnetic field on the speed of sound, 
we use the following definition 
\begin{equation}
v_s^2 = -\frac{\partial P}{\partial E} = \frac{\partial F}{\partial T}\left(\frac{\partial E}{\partial T}\right)^{-1} = \frac{s}{c_v} \, .
\label{vs-def}
\end{equation}
Let us apply the formula (\ref{vs-def}) for the background plus probe system. Expanding at first order in the probe functions, we get:
\beq
v_s^2 = \frac{s_{back}+ s}{c_{v,back}+c_{v}}\,\approx\,{1\over 2}\,-\,
{c_v-2s\over 4 s_{back}}\,\,,
\eeq
where we have taken into account that $c_{v,back}=2 s_{back}$ and, therefore, $v_s^2=1/2$ for the background, as it corresponds to a conformal system in $2+1$ dimensions.  
Since, we can rewrite the ratio ${\cal N}/T \,s_{back}$ in the following form \cite{Jokela:2012dw}
\begin{equation}
{{\cal N}\over T\,s_{back}}\,=\,
{1\over 4}\,{\lambda \over N}\,{q\over b^4}\,\sigma^2\,\,,
\label{N/Ts}
\end{equation}
we arrive at the following expression for the deviation $\delta v_s^2=v_s^2-\frac{1}{2}$:
\bear
&&\delta v_s^2\,\approx\,{\lambda\over  N}\,
{q\,\sigma^2\over 16 \,b^4}\,\,\Bigg[2\,\hat B^2\,\Big({\cal J}(m,\hat B)\,+\,
2\hat B\,{\partial {\cal J}(m,\hat B) \over \partial\hat B}\,\Big)\,+
\rc\rc
&&\qquad\qquad\qquad\qquad
+(3-2b)\,\Bigg(\Big(\,3\,-\,b\,-\,b\,{\partial ( \log c)\over \partial (\log m)}\,\Big)c\,m
-\,4m\,\hat B{\partial c\over \partial\hat B}\Bigg)\Bigg]\,\,, \qquad
\label{delta_vs_general}
\eear
where we have  used  (\ref{cv-s-general})  to compute $c_v-2s$ for the probe.

According to \eqref{def-F} the magnetization of the system is given by the following expression
\be \label{def-mu}
{\cal M} \,  =   \, - \, \left(\frac{\partial F}{\partial B}\right)_T \, = \, - \, \frac{2^{4 \over 3} \pi}{3 \, b} \, {\cal N}_r \, T \, \hat B\,{\cal J}(m,\hat B) \ .
\ee
The magnetic  susceptibility $\chi$ is defined as:
\be  \label{susc}
\chi \, \equiv \, \frac{\partial {\cal M}}{\partial B}\,\,.
\ee

For generic embeddings, which are numerical, also the thermodynamic quantities need to be calculated numerically. However, there are two corners were analytic
results can be obtained. The first one is when we study embeddings with asymptotically large $m$ and the other when the embeddings are massless.  We will consider these two 
cases separately in the next two subsections.

\subsection{Massless embeddings}
For zero mass (and $c=0$) the embedding is necessarily  a black hole embedding and it is more convenient to use the $\theta=\theta(r)$ parameterization. Actually, the massless embeddings in these variables are just characterized by the condition $\theta=\pi/2$. Therefore,  it follows from (\ref{calG_BH}) that the function ${\cal G}(m=0, \hat B)$ is given by the following integral:
\beq
{\cal G}(m=0, \hat B)\,=\,{4b\over r_h^3}
\int_{r_h}^{\infty}dr r^2 \Big(\,
\sqrt{1+
 \Big({r_h\over 2^{{2\over 3}} r}\Big)^{4}\,\hat B^2}\,-\,1
 \Big)\,\,,
\eeq
which can be explicitly performed:
\beq
{\cal G}(m=0, \hat B)\,=\,{4b\over 3}\,\Big[1\,-\,{}_2F_1\Big(-{1\over 2},-{3\over 4},{1\over 4};-{\hat B^2\over 2^{{8\over 3}}} \,\Big)\,\Big]\,\,.
\eeq
Then, it follows that the free energy is given by
\beq
{F\over {\cal N}}\,=\,-1\,+\,{4b\over 3}\,
\Big[1\,-\,{}_2F_1\Big(-{1\over 2},-{3\over 4},{1\over 4};-\Big({3\over 4\pi}\Big)^4\,\,{B^2\over T^4}  \,\Big)\,\Big]\,\,.
\label{F_m0}
\eeq
Let us next compute the entropy density for the massless embeddings. By taking $m=0$ in (\ref{s_general}), we find
\beq
T\,{s(m=0,\hat B)\over {\cal N}} =  - 3\,{\cal G}(m=0,\hat B)\,+\,2\,\hat B^2\,{\cal J}(m=0,\hat B)
\,+\,3\,\,.
\label{s_m0}
\eeq
The integral ${\cal J}(m=0,\hat B)$ can be evaluated explicitly from its definition (\ref{calJ_BH}),
\beq
{\cal J}(m=0,\hat B)\,=\,{b\over 2^{{2\over 3}}}\,
{}_2F_1\Big({1\over 4},{1\over 2},{5\over 4};-{\hat B^2\over 2^{{8\over 3}}} \Big)\,\,.
\label{calJ_m0}
\eeq
Plugging  this result into (\ref{s_m0}), after some calculation,  we arrive at the following simple expression for the entropy density of the  massless embeddings:
\begin{equation}
T \, {s (m=0, B)\over {\cal N}}\,= \, 3 \, - 4 b \, + 4 b \, \sqrt{1 \, + \, \Big({3\over 4\pi}\Big)^4\,\,{B^2\over T^4} } \, .
\label{FiniteTapproxmasslessEntropy}
\end{equation}
Similarly, the internal energy for zero mass is obtained from (\ref{E-calG}):
\beq
{E\over {\cal N}}\Big|_{ m=0}=2+{8b\over 3}\Bigg[
\sqrt{1  +  \Big({3\over 4\pi}\Big)^4\,{B^2\over T^4} }-1+
\Big({3\over 4\pi}\Big)^4\,{B^2\over T^4} \,
{}_2F_1\Big({1\over 2},{1\over 4},{5\over 4};-\Big({3\over 4\pi}\Big)^4\,{B^2\over T^4}\Big)\Bigg]\,\,.
\label{E_massless}
\eeq
We will compute the heat capacity by taking $m=0$ in (\ref{cv-s-general}) and using the remarkable property:
\beq
{\cal J}(m=0, \hat B)\,+\,2\,\hat B {\partial {\cal J}(m=0, \hat B)\over \partial\hat B}\,=\,
{b\over 2^{{2\over 3}}}\,\,{1\over \sqrt{1+{\hat B^2\over 2^{{8\over 3}}}}}\,\,,
\eeq
which combined with (\ref{FiniteTapproxmasslessEntropy}) leads to the simple result:
\beq
T\,{c_v\over {\cal N}}\Big|_{ m=0}=6+8b
\Bigg[{1\over 
\sqrt{1  +  \Big({3\over 4\pi}\Big)^4\,{B^2\over T^4} }}-1\Bigg]\,\,.
\eeq
This result can be confirmed by computing directly the derivative of the internal energy written in (\ref{E_massless}).  The deviation of the speed of sound with respect to the conformal value $v_s^2=1/2$ is readily obtained from (\ref{delta_vs_general}):
\beq
\delta v_s^2\,\approx\,{\lambda\over  2N}\,
{q\,\,\sigma^2\over \,b^3}\,\, 
\Big({3\over 4\pi}\Big)^4\,
{B^2\over T^4\,\sqrt{1  +  \Big({3\over 4\pi}\Big)^4\,{B^2\over T^4} }}\,\,,
\eeq
and the magnetization of the system  at zero mass follows from (\ref{def-mu}) and (\ref{calJ_m0}):
\beq
{\cal M}(m=0, B)\,=\,-{3{\cal N}_r\over 4\pi}\,{B\over T}\,\,
{}_2F_1\Big({1\over 4},{1\over 2},{5\over 4};-\Big({3\over 4\pi}\Big)^4\,\,{B^2\over T^4} \Big)\,\,.
\label{magnetization_m0}
\eeq
We note that the magnetization is always negative and vanishes at zero field.
It is no surprise that the system is diamagnetic. The magnetic field appears with even power inside the DBI action, 
which implies that the spontaneous magnetization vanishes. Moreover,
the DBI action has a specific (plus) sign, meaning that the magnetization is always non-positive.\footnote{In other systems, where the gauge fields have Chern-Simons
terms, their contribution to the magnetization can be positive thus leading to a competition with the DBI part. As a result one might get a positive 
overall magnetization leading to paramagnetism (see \cite{Bergman:2008qv,Jokela:2012vn}), or
even to ferromagnetism (as in \cite{Jokela:2012vn}).}

To obtain the magnetic susceptibility we have to compute the derivative of the right-hand side of (\ref{magnetization_m0}) with respect to the magnetic field (see (\ref{susc})). We get:
\be 
\chi(m=0, B) \,= \, - \frac{3 \, {\cal N}_r}{8 \pi \,T} \,  
\left[{1\over \sqrt{1+\Big({3\over 4\pi}\Big)^4\,{B^2\over T^4} }}\,+\,
{}_2F_1\Big({1\over 4},{1\over 2},{5\over 4};
-\Big({3\over 4\pi}\Big)^4\,{B^2\over T^4}  \Big)\,\right]\,\,.
\ee
In the equations written above for the massless black hole embeddings the dependence on the number of flavors is contained implicitly in the parameter $b$ (see (\ref{b})), while the dependence on the magnetic field  and temperature is manifest. One can take further limits in some of these functions. For example, the $T=0$ values of the free energy (\ref{F_m0}) and entropy (\ref{s_m0}) are:
\begin{equation}
\frac{F(m=0)}{{\cal N}_r}\Big|_{T=0}  \, =  \, \frac{B^{3/2}}{6\sqrt\pi} \, \Gamma\left[\frac{1}{4}\right]^2 \,\,\,,
\qquad\qquad
s (m=0)\Big|_{T=0}  \, = \,  \frac{4 \pi}{3} \, {\cal N}_r \, B \, ,
\label{zeroTapproxmassless_F_s}
\end{equation}
while the magnetization (\ref{magnetization_m0}) of the massless embeddings at zero temperature is:
\begin{equation}
{\cal M} (m=0)\Big|_{T=0} \,= \, - \frac{1}{4 \sqrt{\pi}} \,{\cal N}_r \,  \Gamma\left[\frac{1}{4}\right]^2 \, \sqrt{B} \, . 
\label{zeroTapproxmasslessmagn}
\end{equation}
Let us pause here for a while. We wish to emphasize, that though we were able to produce analytic formulas 
in the special case of massless
embedding, this phase is only relevant for small values of $\hat B$. In particular, the $T=0$ case is never thermodynamically preferred. The phase
diagram will be addressed in Section \ref{sec:catalysis}.

\subsubsection{Small magnetic field}

Let us focus on limits of thermodynamic quantities for the massless case when the magnetic field is small (actually when $B/T^2\to 0$). 
These expressions give the first correction, due to the magnetic field,  to the conformal behavior of the probe at $B/T^2\to 0$. For $F$, $s$, $E$, and $c_v$ we find
\bear
&& {F\over {\cal N}}\,\approx\,-1+2b \,\Big({3\over 4\pi}\Big)^4\,{B^2\over T^4}\,\,,
\qquad\qquad
T\,{s\over {\cal N}}\,\approx\,3+2b \,\Big({3\over 4\pi}\Big)^4\,{B^2\over T^4}\,\,,\rc\rc
&& {E\over {\cal N}}\,\approx\,2\,+\,4b\,\Big({3\over 4\pi}\Big)^4\,{B^2\over T^4}\,\,,
\qquad\qquad
T\,{c_v\over {\cal N}}\,\approx\,6\,-\,4b\,\Big({3\over 4\pi}\Big)^4\,{B^2\over T^4}\,\,.
\eear
Moreover, the variation of the speed of sound at leading order in $B/T^2$ is:
\beq
\delta v_s^2\,\approx\,{\lambda\over  2N}\,
{q\,\,\sigma^2\over \,b^3}\,\, 
\Big({3\over 4\pi}\Big)^4\,
{B^2\over T^4}\,\,,
\qquad\qquad\qquad\qquad
(B/T^2\to 0),
\eeq
and the magnetization becomes:
\beq
{\cal M}\,\approx\,-{3{\cal N}_r\over 4\pi}\,\,{B\over T}\,=\,-
{3 (2-b)\,b^2\sigma \over (4\pi)^2\sqrt{2}}\,N\sqrt{\lambda}\,
{B\over T}\,\,,
\qquad\qquad\qquad
(B/T^2\to 0).
\eeq
It follows that the susceptibility at vanishing magnetic field is:
\beq\label{susc2}
\chi(m=B=0)\,=\,\,-
{3(2-b)b^2\sigma \over (4\pi)^2\sqrt{2}}\,N\sqrt{\lambda}\,
{1\over T}\,\,.
\eeq
Thus, the diamagnetic response of the system goes to zero as the temperature approaches infinity. 
The behavior possessed by \eqref{susc2} closely resembles another $(2+1)$-dimensional construction \cite{Filev:2009ai}. 
In both cases the system behaves as in Curie's law $\chi\propto 1/T$, though they are diamagnetic.  From a dimensional analysis point of view this temperature dependence  is the expected one for the magnetic susceptibility in $2+1$  dimensions, since at high $T$ conformality is restored.

  
\subsection{Approximate expressions for large mass}

When the D6-brane probe remains far away from the horizon, it is possible to obtain analytic results for the free energy and the rest of the thermodynamic quantities.  
Following the analysis of \cite{Jokela:2012dw, Mateos:2007vn}, for large $m$ the embeddings are nearly flat and given by the following expression 
\be
 R(\rho) =R_0+\delta R(\rho) \ ,
\ee
where $R_0$ is a constant and $\delta R(\rho)$ is much smaller than $R_0$. Before calculating the free energy we want an approximate 
expression for the condensate of the theory as a function of the mass and the magnetic field. For this task we need the relationship between $R_0$ and $m$. 
A simple calculation yields the following expansion in powers of $m$
\be\label{R0expansion}
 R_0 = m - a(b)m^{1-\frac{6}{b}} +\frac{1}{4} a_1(b)\hat B^2 m^{1-\frac{4}{b}}+\cdots \ , 
\ee
where the function $a(b)$ is given in equation (B.15) of \cite{Jokela:2012dw}, which we record here for completeness and $a_1(b)$ is
\bea
 a(b) & = &  {3\over 3+2b}\,
\Big[\,{2b\over 3-2b}\,+\,\psi \Big({3\over b}\Big)\,-\,
\psi\Big({3\over 2b}\Big)\,\Big]\\
 a_1(b) & = & \frac{2b}{3-2b}-\psi\left(\frac{3}{2b}\right)+\psi\left(\frac{2}{b}\right) \ ,
\eea 
where $\psi(x)=\Gamma'(x)/\Gamma(x)$ is the digamma function. 
Using \eqref{R0expansion} it is possible to obtain an approximate expression for the condensate as a function of the mass for any number of flavors
\be\label{finiteTcondexpansion}
 c  \, = \, \frac{6\,  b}{4b^2-9} \, m^{-1-{3 \over b}} \, + \, \frac{1}{2} \,\frac{b \, \hat B^2}{3-2b} \, m^{-1-{1 \over b}} \, 
- \, \frac{4}{3} \, \frac{b \, \hat B^2}{3-2b} \, m^{-1-{4 \over b}} \,+  \cdots   \, .
\ee

Using these results in (\ref{calG}) and (\ref{calJ_MN}) it is possible to evaluate the functions  ${\cal G}(m, \hat{B})$ and ${\cal J}(m, \hat{B})$ for large values of the mass parameter $m$. For ${\cal G}$ we get
\begin{equation}
{\cal G}(m, \hat{B})\,=\, 1  + \, 
\frac{b}{2} \,\frac{\hat{B}^2}{m^{\frac{1}{b}}}\, - \, \frac{2 \, b}{ 3\, + 2 b} \, \frac{1}{m^{\frac{3}{b}}} \, - \, 
\frac{b}{3} \,\frac{\hat{B}^2}{m^{\frac{4}{b}}}  \, + \,  \cdots \,\,,
\label{calG_highmass}
\end{equation}
while ${\cal J}$  behaves for large $m$ as:
\begin{equation}
{\cal J}(m, \hat{B})\,=\, b \Bigg[ \, \frac{1 \, }{m^{\frac{1}{b}}}\, - \, 
\frac{2 }{3} \,\frac{1}{m^{\frac{4}{b}}} \Bigg]  \, + \,  \cdots\,\,.
\label{calJ_highmass}
\end{equation}

It is now straightforward to compute the different thermodynamic functions in this high mass regime. Indeed, the free energy and the entropy follow directly by substituting (\ref{calG_highmass}) and (\ref{calJ_highmass}) into (\ref{F-calG}) and (\ref{s_general}), respectively,
\bear
{F\over {\cal N}} & = & - \, \frac{2 \, b}{ 3\, + 2 b} \, \frac{1}{m^{\frac{3}{b}}} \, + \, 
\frac{b}{2} \,\frac{\hat{B}^2}{m^{\frac{1}{b}}} \, - \, \frac{b}{3} \,\frac{\hat{B}^2}{m^{\frac{4}{b}}}  \, \cdots \\
T \, {s\over {\cal N}} & = &  \frac{12 \, b}{ 3\, + 2 b} \, \frac{1}{m^{\frac{3}{b}}} \, + \, 
\, \frac{b \, \hat{B}^2}{m^{\frac{4}{b}}} + \, \cdots \, .
\eear
Similarly, $E$ and $c_v$ can be expanded as
\bear
{E\over {\cal N}} & = &  \frac{b}{2} \,\frac{\hat{B}^2}{m^{\frac{1}{b}}}\, + \,  \frac{10 \, b}{ 3\, + 2 b} \, \frac{1}{m^{\frac{3}{b}}} \, + \, 
\, \frac{2 b}{3} \,\frac{\hat{B}^2}{m^{\frac{4}{b}}} \, + \, \cdots \\
T \, {c_v\over {\cal N}} & = & \frac{60 \, b}{ 3\, + 2 b} \, \frac{1}{m^{\frac{3}{b}}} \, + \, 
\, 2 b\,\frac{\hat{B}^2}{m^{\frac{4}{b}}} \, + \, \cdots \, ,
\eear
and the variation of the speed of sound at large $m$ is given by
\begin{equation}
\delta v_s^2 =  - \, \frac{9}{4} \, {{\lambda\over  N}\, {\sigma^2\over  (3 + 2 b)
(2-b)b^2 }}\,{1 \over m^{3 \over b}} \, + \, \cdots\,\,.
\label{delta_lowT}
\end{equation}
Finally, in the high mass regime the magnetization can be expanded as:
 \begin{equation} 
{\cal M} \, = \, - \, \frac{2^{4 \over 3} \pi}{9} \, {\cal N}_r \, T \, \hat{B} \, \left(\frac{3}{m^{1 \over b}} \, - \, \frac{2}{m^{4 \over b}}\right) \cdots\,\,.
\end{equation}

This regime of large $m$ is achieved when the temperature is low. 
Therefore, it is interesting to compare these results with the ones obtained when $T=0$. We will perform this analysis in the next subsection.

\subsubsection{Zero temperature limit}

One can calculate the different thermodynamic functions at zero temperature by working directly with the parameterization of Section \ref{zeroT_section}, 
in which the embeddings are characterized by the two rescaled  parameters $\tilde m_0$ and $\tilde c_0$ (see (\ref{tilde_m_0}) and (\ref{tilde_c_0})). 
For large values of $\tilde m_0$ one can proceed as above and find an approximate expression of the condensate as a function of the mass:
\beq
 \tilde{c}_0  =  \frac{1}{2} \,\frac{b}{3-2b} \, \tilde m_0^{-1-{1 \over b}} \,+\,\cdots \, ,
 \qquad\qquad
 (\tilde{m}_0\to\infty)\,\,.
 \label{zeroT_tilde_c0_high_mass}
\eeq
Notice that, according to our dictionary (\ref{mu_q_zeroT}), large $ \tilde{m}_0$ corresponds to large $\mu_q^0$ or small $B$. Actually, we can extract the dependence of the condensate on the $B$ field by rewriting (\ref{zeroT_tilde_c0_high_mass}) in terms of the unrescaled parameters 
$c_0$ and $m_0$
\beq
c_0 =  \frac{1}{2} \,\frac{b}{3-2b} \, B^2\,\,m_0^{-1-{1 \over b}} \,+\,\cdots \, . \label{zeroT_c0_high_mass}
\eeq
It is worth pointing out that (\ref{zeroT_tilde_c0_high_mass}) and (\ref{zeroT_c0_high_mass}) can also be obtained by using  (\ref{mc-mc-zero}) and keeping 
the leading terms in the $T\to 0$ limit.  It is instructive to write (\ref{zeroT_c0_high_mass}) in terms of physical quantities. We can use  our dictionary (\ref{mu_q_zeroT}) and (\ref{zeroTdictc}) to translate (\ref{zeroT_c0_high_mass}) to
\beq
 -\frac{\left\langle {\cal O}_q\right\rangle_0}{N}\,=\,
{(2-b)\,b\,\over 8\pi^2\,}\,\,{B_{phys}\over \sqrt{2\lambda}}\,\,
\Bigg[ {\sqrt{\lambda}\over 2\sqrt{2}\,\pi \,\sigma}\,
{B_{phys}\over (\mu_q^0)^2}\,\Bigg]^{{b+1\over 2b}}\,\,,
\eeq
where we have written the result in terms of $B_{phys}\,=\,L^2\,B$. Similarly, by direct calculation or by using the limiting expressions (\ref{mc-mc-zero}),  one finds that the free energy for large $\tilde m_0$ can be approximated as:
\beq
{F\over {\cal N}_r}   \, = \,  \frac{1}{2} \,B^{2}\, m_0^{-{1 \over b}} \,+\, \cdots \, , 
\label{F_Tzero_large_m}
\eeq
which, in terms of physical quantities corresponds to
\beq
F = {(2-b)b^2\,\over 8\pi^2\sqrt{2}}\,{N\over \sqrt{\lambda}}\,
\mu_q^{0}\,B_{phys}\,
\Bigg[ {\sqrt{\lambda}\over 2\sqrt{2}\,\pi \,\sigma}\,
{B_{phys}\over (\mu_q^0)^2}\,\Bigg]^{{b+1\over 2b}}
+\cdots\, .
\eeq
By computing the derivatives  with respect to the magnetic field of the free energy written above one can easily obtain the magnetization and susceptibility at zero temperature in the regime in which $\mu_q^0/\sqrt{B}$ is large.


\section{Magnetic catalysis}\label{sec:catalysis}

In this section we will address the full phase diagram of the system at non-zero magnetic field, zero and non-zero temperature, and in
the presence of background smeared flavors $N_f\ne 0$. There are excellent reviews \cite{Filev:2010pm,Bergman:2012na} which discuss some of the interesting phenomena
that occur on probe-brane systems when an external magnetic field is turned on. To narrow the scope we focus on a particular effect where the magnetic field induces
spontaneous chiral symmetry breaking, known as the magnetic catalysis \cite{Gusynin:1994re,Gusynin:1994xp,Miransky:2002rp}. Only rather recently have we witnessed
attempts in addressing magnetic catalysis that arise away from the probe limit \cite{Filev:2011mt,Erdmenger:2011bw}. 
Current paper constitutes our second step in this direction, the first
one being the construction of a dual to unquenched massive flavors \cite{Bea:2013jxa}. Ultimately, 
one wishes to combine the two and also backreact the magnetic field on the supergravity solution. 
Our goal is very ambitious, but we nevertheless feel that it should not go without serious attempt. In the current paper, we are more modest 
and consider the magnetic field only residing on the probe, but as a background we consider the fully backreacted massless flavored ABJM model.

We will begin our discussion with zero temperature case and then move on to non-zero temperature. Our findings resemble somewhat the results in the
supersymmetric D3-D7 probe brane analysis \cite{Filev:2010pm,Bergman:2012na}. However, surprisingly, the flavor factors go along the ride and we 
can thus analyze the background flavor effects exactly, in contrast to D3-D7 system where the flavors have to be treated perturbatively \cite{Filev:2011mt,Erdmenger:2011bw}. 
We also show that the magnetic catalysis is enhanced (suppressed) with flavor effects at large (small) magnetic field strength when the bare quark mass
is non-zero.

\subsection{Zero temperature}

\begin{figure}[ht]
\center
 \includegraphics[width=0.45\textwidth]{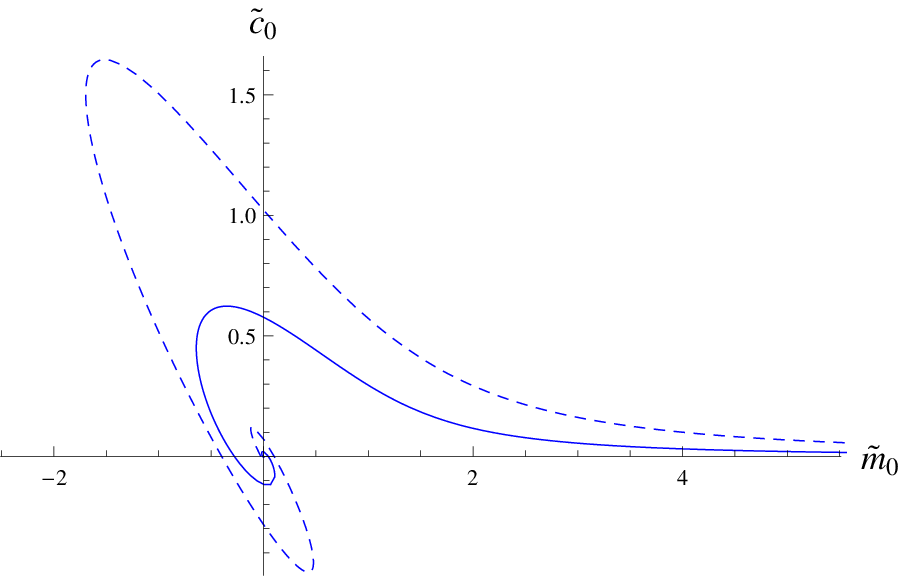}
 \includegraphics[width=0.45\textwidth]{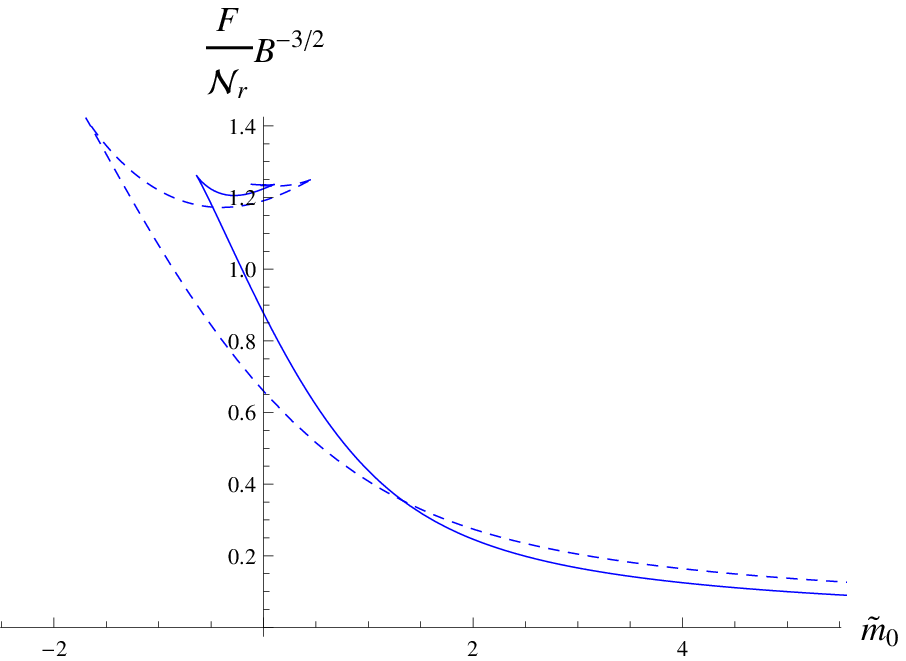}
 \caption{Plot of the condensate $\tilde c_0$ (left) and the free energy (right) versus the rescaled mass $\tilde m_0$. 
 The solid blue is $\hat\epsilon=0$ and the dashed blue is $\hat\epsilon=\infty$. 
 Both of the curves on the right panel start at the value (at $\tilde m_0=0$) $\frac{\Gamma(1/4)^2}{6\sqrt\pi}$ as extracted from (\ref{zeroTapproxmassless_F_s}). 
 Notice that there is no phase transition, since for all $\tilde m_0\ge 0$ we reside on the same solution. 
 }
 \label{fig:FforT0}
\end{figure}

Consider the equation of motion for the embedding $\tilde R=\tilde R(\tilde\rho)$ as derived from the action (\ref{eq:zeroTaction}) at zero temperature $T=0$.
At non-zero $B$ the supersymmetry is broken, and thus the D6-brane has a profile (as in (\ref{eq:zeroTasymp})) which 
depends on $\tilde\rho$. For large $\tilde m_0$, the condensate can be obtained analytically (see (\ref{zeroT_tilde_c0_high_mass})), but for small values of $\tilde m_0$, the $\tilde c_0$ needs to be numerically solved
for. In Fig.~\ref{fig:FforT0} (left panel) we display the parametric plot of $\tilde c_0$ versus $\tilde m_0$, which was generated by 
varying the IR value $\tilde R(0)=\tilde R_0$
and shooting towards the $AdS$ boundary.\footnote{Notice that only $\tilde m_0\ge 0$ embeddings are physical; otherwise the angle $\theta>\pi/2$.} 
In Fig.~\ref{fig:FforT0} (right panel) we plot the free energy as 
a function of $\tilde m_0$. We find several possible solutions for some given small $\tilde m_0$, but immediately infer that the physical solution is the one with 
lowest free energy. This corresponds to the solution with larger condensate, {\emph{i.e.}}, corresponding to the right arms of $(\tilde m_0,\tilde c_0)$ curves.
We also note, that the large $\tilde m_0$ tail corresponds to the analytic behavior (\ref{zeroT_tilde_c0_high_mass}), whereas zooming in toward origin of the plot 
would probably result in a self-similar behavior of the equation of state, similarly as was analyzed in \cite{Jokela:2012dw}.

\subsubsection{Zero bare mass}

An important result is that the $\tilde m_0=0$  embedding has a non-zero fermion condensate $\tilde c_0\propto \langle {\cal O}_q\rangle_0$. 
In terms of physical quantities, the relation (\ref{zeroTdictc}) between 
$ \langle {\cal O}_q\rangle_0$ and $\tilde c_0$ has  been written in the appendix (eq. (\ref{Oq_zero-czero_Bphys})).  
The introduction of the magnetic field $B$ has therefore induced a spontaneous chiral symmetry breaking.


\begin{figure}[ht]
\center
 \includegraphics[width=0.7\textwidth]{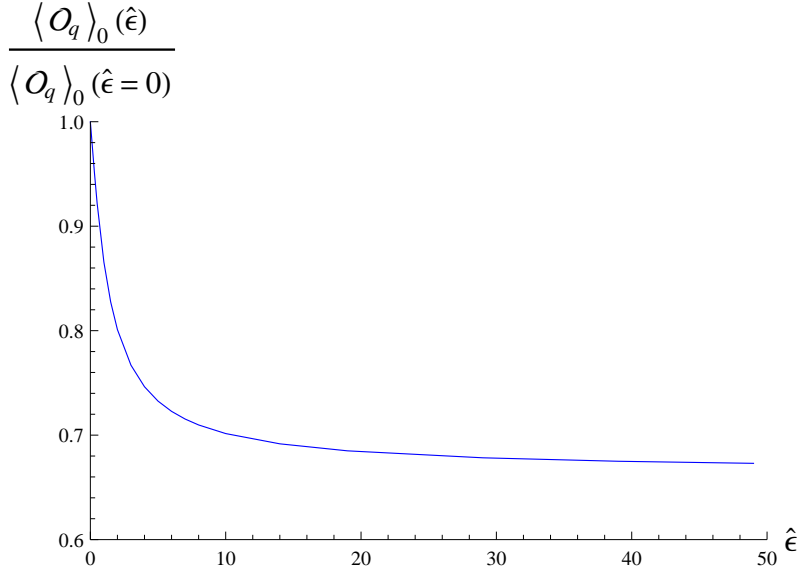}
 \caption{Plot of the condensate against the deformation parameter at zero bare mass. At infinite flavor the condensate reaches a constant value for fixed physical magnetic 
 field $B_{phys}$. Notice that we normalized the depicted quantity to unity
in the quenched limit. }
 \label{fig:condvseps_withB_zeroT}
\end{figure}

We now focus on the flavor effects. While we find in Fig.~\ref{fig:FforT0} that the $\tilde c_0$ grows with increasing $\hat\epsilon$, the condensate $\langle {\cal O}_q\rangle_0$
given in (\ref{Oq_zero-czero_Bphys}) actually has the opposite behavior with more flavor.  In Fig.~\ref{fig:condvseps_withB_zeroT} we plot essentially $\langle {\cal O}_q\rangle_0$ against $\hat\epsilon$ and
see that the condensate actually decreases monotonously with $\hat\epsilon$ for fixed $B_{phys}$.  In a different system \cite{Filev:2011mt} the tendency for the condensate to decrease with flavor was also observed. 
At infinite flavor $\hat\epsilon\to\infty$, $\langle {\cal O}_q\rangle_0$ reaches a constant non-zero value.

\subsubsection{Non-zero bare mass}\label{sec:suppress}

While the magnetic catalysis is the main focus of this paper, it is interesting to study the case with non-vanishing bare mass. In other words, we wish to study
the system when the chiral symmetry is {\emph{explicitly}} broken, rather than {\emph{spontaneously}}, and ask what does the condensate care about the magnetic field and 
background flavor. 

When the bare mass is non-zero it is convenient to study the value of the condensate for a fixed value of $\mu_q^0$. From our dictionary (\ref{mu_q_zeroT}) we can relate the mass parameter $\tilde m_0$ to $\mu_q^0$ and the physical magnetic field. We get:
\beq
\tilde m_0\,=\,\sqrt{2\sqrt{2}\,\pi\,\sigma}\,\,
\Big[{(\mu_q^0)^2\over \sqrt{\lambda}\,B_{phys}}\Big]^{{1\over 2}}\,\,.
\label{m0_Bphys}
\eeq
The formula (\ref{m0_Bphys}) enables us to plot the condensate against the magnetic field itself, in units of $\mu_q^0$  rather than the rescaled mass $\tilde m_0$. Indeed, let us now consider the quantity:
\beq
-{\lambda\over (\mu_q^0)^2}\,\,
{\left\langle {\cal O}_q\right\rangle_0\over N}\,=\, 
{(3-2b)(2-b)\over 4\pi^2\sqrt{2}}\,\,
 {\sqrt{\lambda}\,B_{phys}\over (\mu_q^0)^2}\,\,\tilde c_0\,\,.
 \label{vev_bare_mass}
 \eeq
The condensate parameter $\tilde c_0$ depends non-trivially on $\tilde m_0$ (the precise dependence must be found by numerical calculations), which in turn can be written as in (\ref{m0_Bphys}). Thus, it follows that the left-hand side  of 
(\ref{vev_bare_mass}) depends on $\sqrt{\lambda}\,B_{phys}/ (\mu_q^0)^2$. 

\begin{figure}[ht]
\center
 \includegraphics[width=0.45\textwidth]{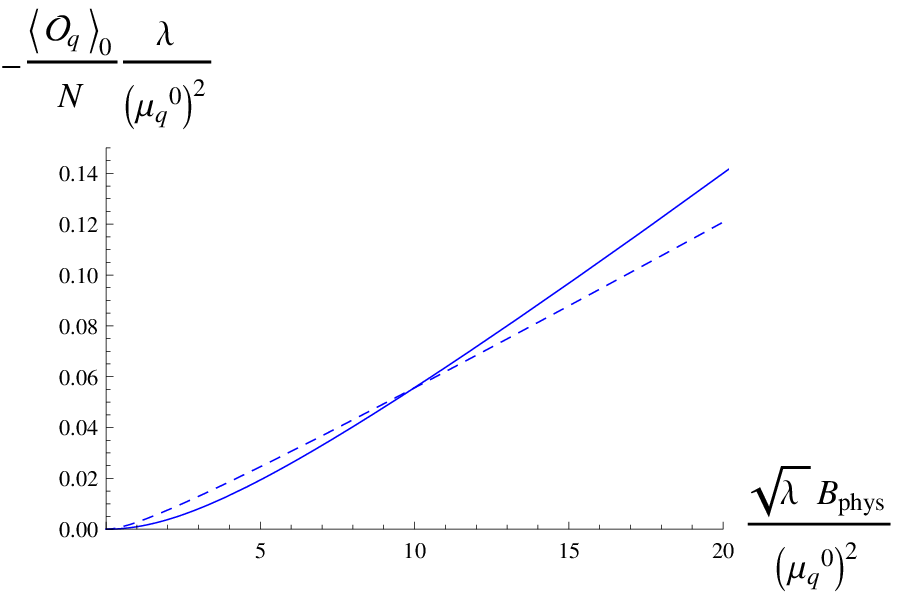}
 \qquad
  \includegraphics[width=0.45\textwidth]{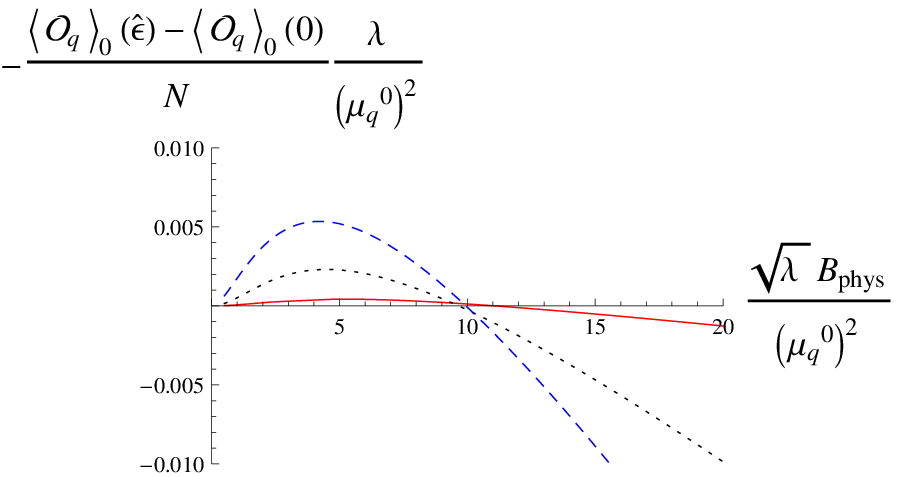}
\caption{Plots of the condensate versus the magnetic field, when the bare quark mass is non-zero. 
The solid blue is $\hat\epsilon = 0$ and the dashed blue is $\hat\epsilon = 10$. In the right panel we have also included the $\hat\epsilon=0.1$ (solid red)
and the $\hat\epsilon=1$ (dotted black) curves.}
 \label{fig:condvsB_m}
\end{figure}

In Fig.~\ref{fig:condvsB_m} we show the condensate against the magnetic field and find that it increases monotonously.
Moreover, for small $B_{phys}/(\mu_q^0)^2$ the condensate for the flavored theory is larger than the unflavored one. 
Thus, for small $B_{phys}$ or large $\mu_q^0$, the flavors produce an enhancement of the chiral symmetry breaking. 
In Fig.~\ref{fig:condvsB_m} we illustrate this flavor effect by plotting the difference of condensates as a function of $B_{phys}/(\mu_q^0)^2$. Actually, for small values of $B_{phys}/(\mu_q^0)^2$ we can use the approximate expression 
(\ref{zeroT_tilde_c0_high_mass}) to estimate $\tilde c_0$.
Plugging (\ref{zeroT_tilde_c0_high_mass}) and (\ref{m0_Bphys}) into (\ref{vev_bare_mass}) we get for small $B_{phys}/(\mu_q^0)^2$
\beq
-{\lambda\over (\mu_q^0)^2}\,\,
{\left\langle {\cal O}_q\right\rangle_0\over N}\,\approx\, 
{(2-b)b\over 8\sqrt{2}\,\pi^2}\,\,
(2\pi\,\sqrt{2}\,\sigma)^{-{1\over 2}-{1\over 2b}}\,\,
\Bigg[ {\sqrt{\lambda}\,B_{phys}\over (\mu_q^0)^2}\Bigg]^{{3\over 2}+{1\over 2b}}\,\,.
\eeq
Thus, we get a power law behavior with an exponent which depends on the number of flavors and matches the numerical results. 

For large values of $B_{phys}/(\mu_q^0)^2$ the flavors suppress the condensate and we have a behavior similar to the massless case. 
Curiously, this change of behavior occurs at values of $B_{phys}/(\mu_q^0)^2$ which 
are almost independent of the number of flavors (see Fig.~\ref{fig:condvsB_m}). 


\subsection{Non-zero temperature}

Having understood the basic physics behind introducing the magnetic field, let us now heat up the system and study what happens. In addition to the MN embeddings, we now
also have the BH embeddings at our disposal. At zero magnetic field, $B=0$, we recall that there is going to be a phase transition from the MN phase to the BH phase as the temperature
is increased \cite{Jokela:2012dw}. The black hole begins to increasingly attract the probe D6-brane. Turning on $B$ has the opposite effect, in some sense
the magnetic field makes the D6-brane to repel. We thus have two competing effects in play and we need to explore the four-dimensional phase space $(T,B,\mu_q,\hat\epsilon)$,
to find out which phase is thermodynamically preferred.
Recall that at non-zero temperature we can form the dimensionless ratio (\ref{eq:Bhat}), 
which for the physical magnetic field is   $\hat B_{phys} \equiv 2^{4/3}\frac{B_{phys}}{r_h^2}$ and that the bare mass $\mu_q$ was introduced in \cite{Jokela:2012dw}, see below. This narrows down the phase space down to 
three dimensions $(\hat B_{phys},\mu_q,\hat\epsilon)$. Let us begin our journey in the simpler case with vanishing bare quark mass $\mu_q=0$.

\subsubsection{Zero bare mass}

\begin{figure}[ht]
\center
 \includegraphics[width=0.45\textwidth]{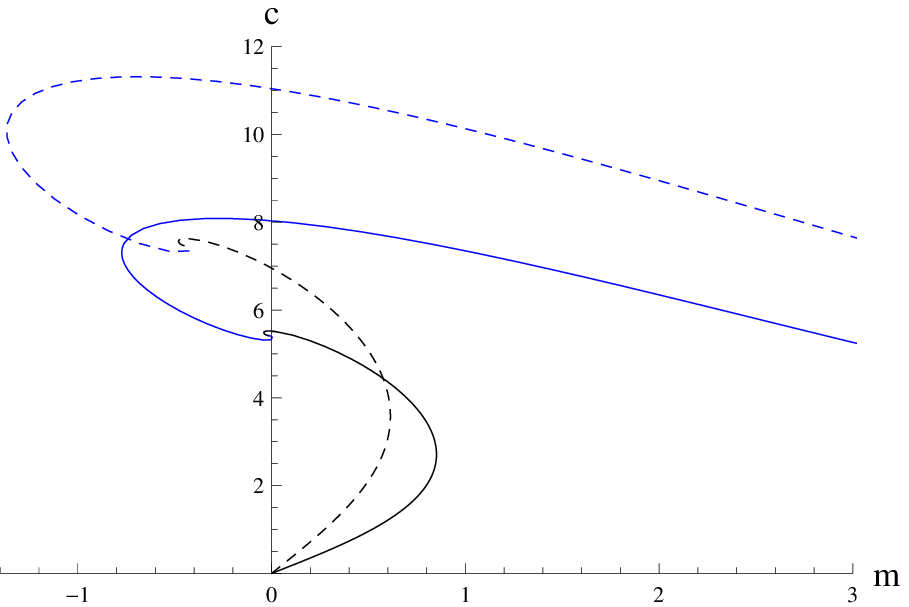}
 \includegraphics[width=0.45\textwidth]{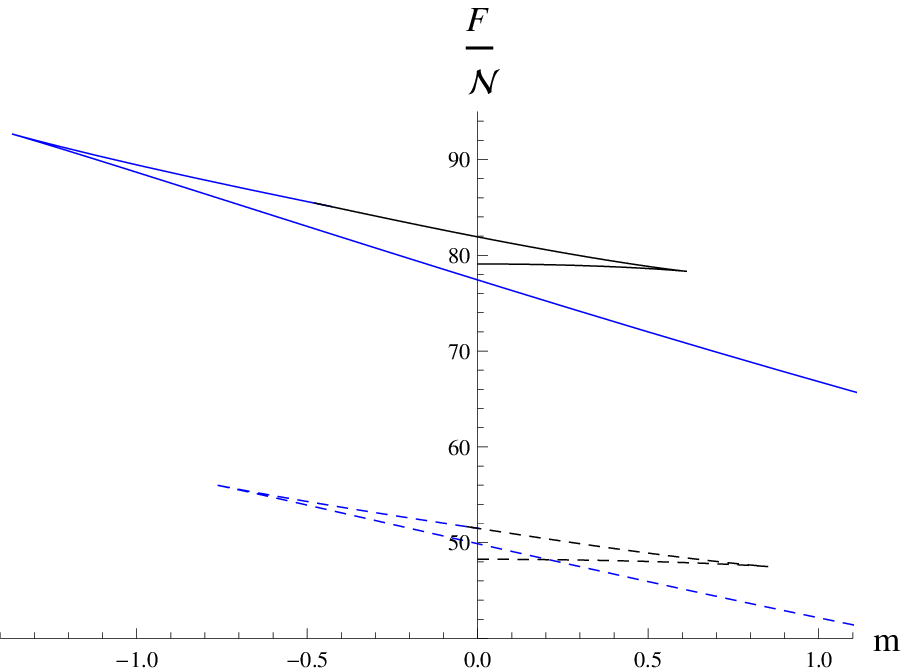}
 \caption{Plot of the condensate $c$ (left) and the free energy (right) versus $m$ at $\hat\epsilon = 0$. 
 The solid curves are for $\hat B=20$ and the dashed curves are for $\hat B=15$. The blue color stands for MN embeddings and black for BH embeddings.
 The phase transition for $m=0$ is between these two cases, around $\hat B_{crit}(\hat\epsilon=0)\sim 17.8$, above this critical value the chiral symmetry is always broken. 
 }
 \label{fig:FforTne0}
\end{figure}

\begin{figure}[ht]
\center
 \includegraphics[width=0.45\textwidth]{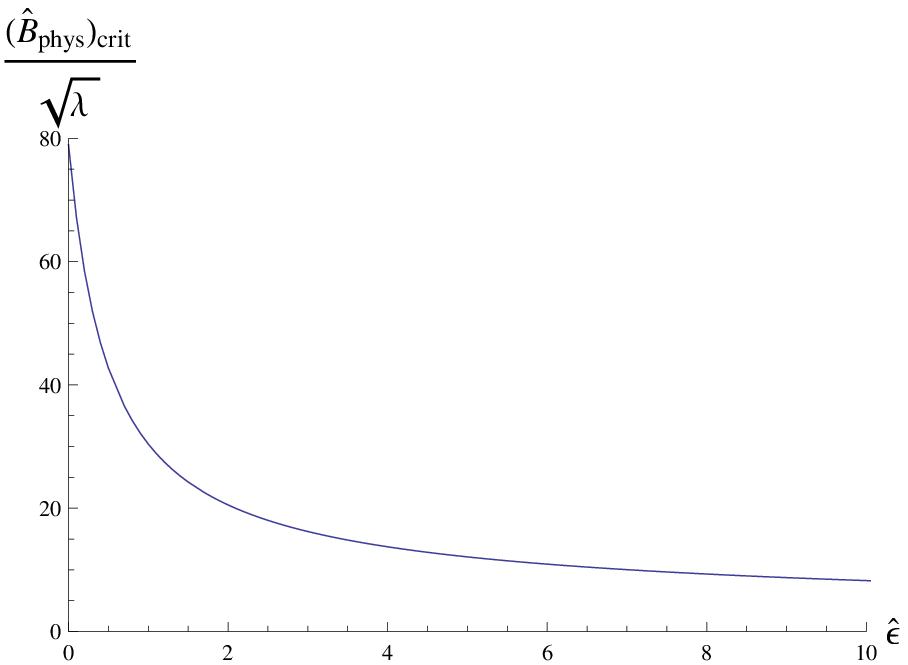}
 \includegraphics[width=0.45\textwidth]{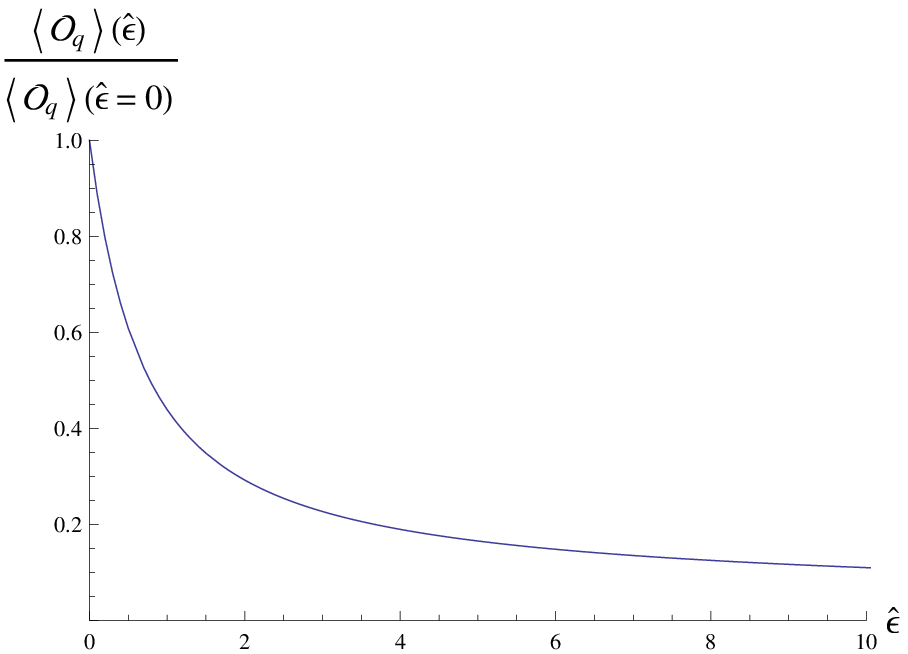}
 \caption{On the left we plot the phase diagram for $m=0$ in the $(\hat\epsilon,\hat B)$-plane. Above the curve, the chiral symmetry is spontaneously broken whereas below the curve the  condensate is zero. We note that the critical magnetic field needed, at fixed temperature, to break the chiral symmetry decreases with the number of flavors, leading to the decrease
 of the condensate (and asymptotically vanishing due to the screening function $\sigma$), as depicted on the right. 
 }
 \label{fig:critvseps}
\end{figure}

We start exploring the phase space in the case where we set $\mu_q=0$. This slice of the full phase diagram is easily obtained. At any given $\hat\epsilon$ we only have two options,
either the system is in the chirally symmetric BH phase (small $\hat B_{phys}$) or the system is in the MN phase (large $\hat B_{phys}$) and
the chiral symmetry is broken; see Fig.~\ref{fig:FforTne0}. There is a first order 
phase transition at some critical $\hat B^{phys}_{crit}$, which depends on $\hat\epsilon$.
Above this critical $\hat B^{phys}_{crit}$, the BH phase is never reached and thus the chiral symmetry is spontaneously broken. 
The phase diagram $(\hat\epsilon,\hat B)$ is presented in Fig.~\ref{fig:critvseps} (left panel). The curve plotted $\hat B^{phys}_{crit} = \hat B^{phys}_{crit}(\hat\epsilon)$ shows 
that the critical magnetic field decreases with increasing number of flavors. In other words, at fixed temperature, the more
flavor there is the smaller magnetic field is needed to realize magnetic catalysis. As a consequence the critical condensate will also be smaller with more flavors, as is
visible in Fig.~\ref{fig:critvseps} (right panel).

\begin{figure}[ht]
\center
 \includegraphics[width=0.65\textwidth]{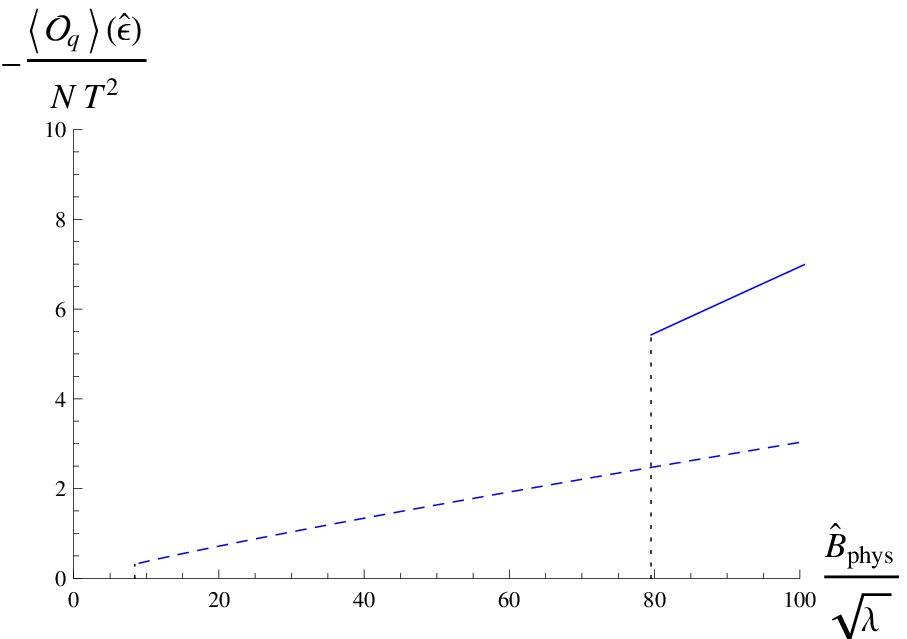}
 \caption{The condensate versus the magnetic field $\hat B$ at $\mu_q = 0$. The solid blue curve is for $\hat\epsilon=0$ and the dashed blue curve is for $\hat\epsilon=10$. 
 On the left of the curves the condensate is zero. Notice
 that at critical magnetic field $\hat B_{crit}(\hat\epsilon)$ there is a first order phase transition (from the chirally symmetric BH phase) where the condensate jumps to a non-zero value, thus acting as an
 order parameter for the transition (to the chirally broken MN phase).}
 \label{fig:order_m0}
\end{figure}

We finish this subsection by presenting the graph Fig.~\ref{fig:order_m0}, which represents the condensate as a function of the magnetic field at selected flavor deformation 
parameters ($\hat\epsilon=0$ and $10$) and zero bare quark mass. The swallow-tail structures of the free energy graphs are indications of the 
first order phase transition, and from Fig.~\ref{fig:order_m0} we conclude that the condensate acts as an order 
parameter: at critical $\hat B_{crit}^{phys}$ the condensate jumps to a non-zero value and increases thereafter.
From the numerics we also infer, that for large $\hat B$,
\be
 \frac{\langle {\cal{O}}_q\rangle}{N} \sim T^2 \hat B^{(3-b)/2} \sim B\cdot\left(\frac{T}{\sqrt B}\right)^{\gamma_m} \quad \ , \hat B\gg 1 \ .
\ee
This behavior conforms with the $T=0$ result (\ref{zeroTdictc}) at $m=0$ (recall the relation (\ref{eq:dictrelation})).

\subsubsection{Non-zero bare mass}

To complete the investigation of the phase diagram, let us turn on a non-zero bare mass at non-zero temperature. 
Recall that the bare mass $\mu_q$ is given by \cite{Jokela:2012dw}
\be\label{eq:muq}
 \frac{\mu_q}{T\sqrt\lambda} = \frac{2^{5/6}\pi}{3}m \ .
\ee
Given the relation (\ref{eq:muq}), instead of directly fixing the bare mass to some value, we can fix $m$ (for any flavor deformation parameter $\hat\epsilon$). We just
need to keep in mind that larger $m$
will then correspond to smaller temperatures, and vice versa.

\begin{figure}[ht]
\center
 \includegraphics[width=0.65\textwidth]{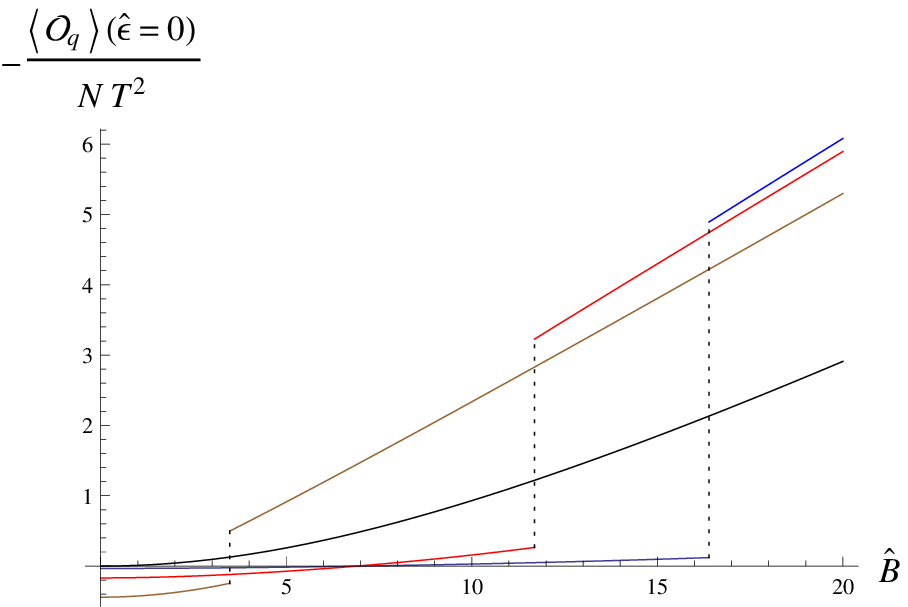}
 \caption{The condensate versus the magnetic field $\hat B$ at various fixed bare masses at quenched case $\hat\epsilon=0$. The critical magnetic field $\hat B_{crit}$, whose 
 values correspond to the dotted vertical line segments,
 decreases as $m$ increases, and the respective curves read $m=0.1$ (blue), $m=0.5$ (red), $m=1$ (brown), and $m=5$ (continuous black).}
 \label{fig:condvsBhat_mfixed}
\end{figure}

We anticipate that there are essentially two different cases, depending on whether $m$ is small or large. 
In Fig.~\ref{fig:condvsBhat_mfixed} we depict the condensate as a function of $\hat B$ for various $m$ at $\hat\epsilon =0$; the $\hat\epsilon >0$ is qualitatively the same with smaller
$\hat B_{crit}$'s.
We find that for any given $\hat\epsilon$ there exists a large
enough $m$ such that the system is always in the chirally broken MN phase for any $\hat B$. For small values of $m$, there can be a phase transition from the chirally symmetric
BH phase to a broken MN phase at some critical $\hat B_{crit}$.


\section{Conclusions}
\label{conclusion}

Let us shortly recap the main results of our work. We studied the ABJM Chern-Simons matter model with dynamical flavors, added as smeared flavor D6-branes.  
Our black hole geometry includes the backreaction of dynamical massless flavors at fully non-linear order in the flavor deformation parameter (\ref{hat-epsilon_def}). 
We investigated the effect of the inclusion of an external magnetic field on the worldvolume of an (additional) probe D6-brane and focused on the 
flavor effects from the smeared D6-branes of the background. We obtained the different 
thermodynamic functions for the probe and explored the corresponding phase diagram. In some corners of this phase space we were able to obtain analytic results.

At zero temperature, for any magnetic field strength, the system was always 
in the chirally broken MN phase; a phenomenon called magnetic catalysis. 
At large (small) magnetic field strength, at non-vanishing bare quark mass, the condensate was found decreasing (increasing) with the number of flavors. 
In other words, for small masses the magnetic catalysis is suppressed whereas for large values of the mass it is enhanced given more flavors in the background.
This behavior could morally be thought of as inverse magnetic catalysis in the sense of \cite{Bali:2012zg,Preis:2010cq}, although is technically different.

At non-zero temperature there was a critical magnetic field above which the magnetic catalysis took place. The condensate acted as an order parameter for the first order
phase transition between the transition from the chirally symmetric BH phase to the broken MN phase.
We found that the critical magnetic field was smaller for more flavors, which we interpret as an enhancement of the magnetic catalysis.

Let us finally discuss some possible extensions of our work.  First of all, we could analyze the effects of having unquenched {\it massive}  flavors. 
The corresponding background for the ABJM theory at zero temperature has been recently constructed in \cite{Bea:2013jxa}. 
It would be interesting to explore in this setup how the flavor effects on the condensate are enhanced or suppressed as the mass of the unquenched 
quarks is varied, and to compare with the results found here in Section \ref{sec:suppress}. 
Also, one could try to include the effect of the magnetic field on the unquenched quarks. 
For this purpose a new non-supersymmetric background must be constructed  first (see \cite{Filev:2011mt,Erdmenger:2011bw} for a similar analysis in the D3-D7 setup). 
To complete the phase structure of the model we must explore it at non-zero chemical potential. 
This would require introducing a non-vanishing charge density by exciting  additional components of the worldvolume gauge field. 

\vspace{0.2cm}

{\bf \large Acknowledgments}
We thank Yago Bea and Johanna Erdmenger for discussions and Javier Mas for collaboration at the initial stages of this work. We are specially grateful to 
Veselin Filev for his comments and help. 
N.~J. and A.~V.~R. are funded by the Spanish grant FPA2011-22594, by the Consolider-Ingenio 2010 Programme CPAN (CSD2007-00042), by Xunta de
Galicia (Conselleria de Educaci\'on, grant INCITE09-206-121-PR and grant PGIDIT10PXIB206075PR), and by FEDER. N.~J. is
also supported by the Juan de la Cierva program.  
D.~Z. is funded by the FCT fellowship SFRH/BPD/62888/2009. Centro de F\'isica do Porto is partially funded by FCT through the projects CERN/FP/116358/2010 and PTDC/FIS/099293/2008.

\newpage

\appendix
\vskip 1cm
\renewcommand{\theequation}{\rm{A}.\arabic{equation}}
\setcounter{equation}{0}

\section{Zero temperature dictionary}\label{appendix}

The relation between the mass and the parameter $\tilde m_0$ has been worked out in detail in Section \ref{runnning_mass_section}.  
In this Appendix  we work out the dictionary for  the condensate at $T=0$, which we will denote by  $\left\langle {\cal O}_q\right\rangle_0$.
A similar analysis at non-zero temperature was 
presented in appendix D of \cite{Jokela:2012dw}. For simplicity, in this Appendix we use units in which $\alpha'=1$.

Let $\mu_q^0$ be the bare quark mass at zero temperature, 
whose explicit expression in terms of $\tilde m_0$ and $B$ has been derived in Section \ref{runnning_mass_section} (Eq.~(\ref{mu_q_zeroT})).  
The expectation value  $\left\langle {\cal O}_q\right\rangle_0$ is obtained as the derivative with respect to $\mu_q^0$ of the zero temperature free energy:
\beq
\left\langle {\cal O}_q\right\rangle_0\,=\,{\partial F\over \partial \mu_q^0}\,\,.
\label{VEV_def}
\eeq
To compute the derivative  in (\ref{VEV_def}) we apply the chain rule:
\beq
{\partial F\over \partial \mu_q^0}\,=\,
{\partial F\over \partial \tilde m_0}\,{\partial  \tilde m_0\over \partial \mu_q^0}\,\,,
\eeq
and use \cite{Jokela:2012dw}:
\beq
{\partial F\over \partial \tilde m_0}\,=\,-{3-2b\over b^2}\,B^{{3\over 2}}\,\tilde c_0\,
{\cal N}_r\,\,,
\eeq
where $\tilde c_0\,=\,B^{{b-3\over 2}}\,c_0$. We get:
\beq
\left\langle {\cal O}_q\right\rangle_0\,=\,-{3-2b\over b^2}\,B^{{3\over 2}}\,\tilde c_0\,\,
{\tilde m_0\over  \mu_q^0}\,\,{\cal N}_r\,\,.
\eeq
Using (\ref{mu_q_zeroT}) and  the expression of ${\cal N}_r$, we find:
\beq
{\tilde m_0\over  \mu_q^0}\,\,{\cal N}_r
\,=\,{(2-b)\,b^2\,\sigma\over 4\pi}\,N\,B^{-{1\over 2}}\,\,.
\eeq
Therefore, we have the following relation between $\left\langle {\cal O}_q\right\rangle_0$ and $\tilde c_0$:
\beq
-{\left\langle {\cal O}_q\right\rangle_0\over N}\,=\, 
{(3-2b)(2-b)\sigma\over 4\pi}\, B\,\tilde c_0\,\,,
\label{Oq_zero-czero}
\eeq
which, after including the appropriate power of $\alpha'$,  coincides with the expression written in (\ref{zeroTdictc}). 
Let us finally write (\ref{Oq_zero-czero}) in terms of the physical magnetic field $B_{phys}$ given by (\ref{Bphys_def}). We find
\beq
-{\left\langle {\cal O}_q\right\rangle_0\over N}\,=\, 
{(3-2b)(2-b)\over 4\pi^2\sqrt{2}}\,\, {B_{phys}\over \sqrt{\lambda}}\,\,\tilde c_0\,\,.
\label{Oq_zero-czero_Bphys}
\eeq

\end{document}